\documentclass[twocolumn,superscriptaddress,longbibliography,10pt]{revtex4-1}
\usepackage{amsmath,amssymb,amsfonts,mathtools}
\usepackage{epsfig}

\usepackage{slashed}
\usepackage{color}

\begin{document}

\title{Computing Light-Front Wave Functions Without Light-Front Quantization: \\ A Large-Momentum Effective Theory Approach}

\author{Xiangdong Ji}
\email{xji@umd.edu}
\affiliation{Center for Nuclear Femtography, SURA, 1201 New York Avenue. NW, Washington, DC 20005, USA}
\affiliation{Maryland Center for Fundamental Physics,
Department of Physics, University of Maryland,
College Park, Maryland 20742, USA}

\author{Yizhuang Liu}
\email{yizhuang.liu@uj.edu.pl}
\affiliation{Institute of Theoretical Physics, Jagiellonian University, 30-348 Kraków, Poland}

\date{\today}

\begin{abstract}
Light-front wave functions play a fundamental role in the light-front quantization approach to QCD and hadron structure. However, a naive implementation of the light-front quantization suffers from various subtleties including the well-known zero-mode problem, the associated rapidity divergences which mixes ultra-violet divergences with infrared physics, as well as breaking of spatial rotational symmetry. We advocate that the light-front quantization should be viewed as an effective theory in which small $k^+$ modes have been effectively ``integrated out'', with an infinite number of renormalization constants. Instead of solving light-front quantized field theories directly, we make the large momentum expansion of the equal-time Euclidean correlation functions in instant quantization as an effective way to systematically calculate light-front correlations, including the light-front wave function amplitudes.  This large-momentum effective theory accomplishes an effective light-front quantization through lattice QCD calculations. We demonstrate our approach using an example of a pseudo-scalar meson wave function.

\end{abstract}

\maketitle

\section{Introduction}

Light-front (LF) quantization (LFQ) or formalism is a natural language for parton physics
in which partons are made manifest in all stages of calculations. The goal of the
Hamilton formulation of LFQ is to solve non-perturbative quantum chromodynamics (QCD) just like a non-relativistic quantum
mechanical system, i.e., diagonalizing the Hamiltonian and obtaining the wave-functions for
the QCD bound states
~\cite{Brodsky:1997de},
\begin{equation}\label{eq:LFeigen}
      \hat P^-|\Psi_n\rangle = \frac{M^2_n}{2P^+}|\Psi_n\rangle \,
\end{equation}
where $P^+$ is the LF momentum (see Eq. (5) below for light-front coordinate definition) and $\hat P^-$ is the LF Hamiltonian, and $M_n$ and $|\Psi_n\rangle$
are a hadron mass and wave function, respectively.
Then all the partonic densities and correlations functions may be calculated as the expectation values of the LF wave-functions (LFWF). Moreover, like in condensed matter physics, knowing quantum many-body wave-functions allows one to understand interesting aspects of quantum coherence and entanglement, as well as the fundamental nature of quantum systems. Therefore, a practical realization of LFQ program clearly would be a big step forward in understanding the fundamental structure of hadrons, particularly the nucleon (proton and neutron), which are the fundamental building blocks of visible matter.

To be sure, wave-functions for the hadron bound states are not the most natural objects in quantum field theory (QFT) due to the non-trivial QCD vacuum, ultra-violet (UV) divergences as well as Lorentz symmetry, according to the latter the space and
time shall be treated on the equal footing. The proton and other hadrons are excitations of the QCD vacuum which by themselves are very complicated because of the well-known phenomena of chiral symmetry breaking and color confinement.
To build a proton on the top of this vacuum, one naturally has a question about what part of the
wave-function reflects the property of the bound state and what reflects the QCD vacuum:
It is the difference which yields the properties
of the proton that are experimentally measurable. There is no clean way to make this separation unless one builds the proton out of elementary excitations or quasi-particles that
do not exist in the vacuum, as often have been done in condensed matter systems.

The parton degrees of freedom in the infinite momentum frame (IMF) fulfill the above purpose to some degree. Due to the kinematic effects, all partons in
the vacuum have longitudinal momentum $k^+=0$, and to some degree of accuracy
(incorrect for higher-twist observables as we will discuss),
the proton is made of partons with $k^+\ne 0$. This separation of
degrees of freedom is particularly welcome, making
a wave-function description of the proton more natural and interesting
than in any other frame. Moreover, one can impose
a infrared (IR) cut-off on the $k^+\ge \epsilon$ in the effective Hilbert space,
such that all physics below $k^+=\epsilon$ are taken into account through
renormalization procedures. The result is an effective LF theory with ``trivial''
vacuum,
\begin{equation}
     a_{k\lambda}|0\rangle = b_{p\sigma}|0\rangle = d_{p\sigma}|0\rangle=0 \ .
\end{equation}
where $|0\rangle $ is the vacuum of QCD, $a_{k\lambda}$ is an annihilation operator for a gluon with momentum $k$ and polarization $\lambda$, and similarly the annihilation operators  $b_{p\sigma}$ and $d_{p\sigma}$ for quark and anti-quarks.
Therefore schematically one can write down the Fock-space expansion for the proton state in LF gauge $A^+=0$,
  \begin{align}
|P\rangle=\sum_{n=1}^{\infty} \int d\Gamma_n \psi_n(x_i,\vec{k}_{i\perp})\prod a^{\dagger}_i(x_i, \vec{k}_{i\perp})|0\rangle \ .
\end{align}
where we use the generic notation $a^\dagger$ to denote quarks and gluon quanta on the LF, the phase-space integral reads $d\Gamma_n=\prod \frac{dk^+_id^2k_{i\perp}}{2k^+_i(2\pi)^3}$, with $x_i = k_i^+/P^+$,
$\psi_n(x_i,\vec{k}_{i\perp})$
are LFWF or LFWF amplitudes or simply LF amplitudes.
The latter are a complete set of non-perturbative quantities which describe the partonic landscape of the proton.

The nature of QCD vacuum in LFQ has been continuously debated in the
literature. One knows a priori that in relativistic QFT, the vacuum state is
boost-invariant and frame independent, and hence it shall be independent of quantization formalism. In fact, it have been proved in Ref.~\cite{Nakanishi:1976yx,Nakanishi:1976vf}
that not only the non-trivial QFT vacuum can not be simple, not all Green's functions of the theory can not
poses generic meaningful restrictions to the null-planes $\xi^+=$ constant.  Therefore,
zero modes do contain non-trivial dynamics and contribute to the
high-twist properties of the proton, such as mass and transverse spin~\cite{Ji:2020baz}. As we explain in the next section,
an effective theory view of LFQ simply cuts off the zero-mode complication
and relegates these physics to renormalization constants, although in some cases, the zero-modes can be treated explicitly in the literature ~\cite{Heinzl:1991myy,Yamawaki:1998cy}.

With the above caveat, it is possible to invert the LF quantum state, by
express the LFWFs in terms of invariant matrix element amplitudes,
\begin{align}\label{eq:LFcorre}
\psi_n(x_i,\vec{k}_{i\perp})=\langle 0|\prod a_i(x_i,\vec{k}_{i\perp})|P\rangle \ .
\end{align}

By properly restoring gauge-invariance through LF gauge links $\exp(ig\int A^+d\lambda)$ and imposing regularizations, the above
amplitudes can be calculated not only in LF theory but can also be accessed through the large momentum effective theory (LaMET)
approach~\cite{Ji:2013dva,Ji:2014gla,Cichy:2018mum,Ji:2020ect}. Therefore, one can actually obtain a LF picture of the proton without going through the explicit LFQ, or one can effectively obtain the results of
LFQ through instant quantization at a large momentum frame.

The LFWFs also naturally arise in high-energy processes.  For exclusive processes~\cite{Lepage:1979zb,Lepage:1980fj}, one usually probe a given Fock component and the non-perturbative distributions for these components are exactly LFWFs and the $\vec{k}_\perp$ integrated
LFWF or the Distribution Amplitude (DA). They have been applied to various form
factors such as the pion electro-magnetic form factor~\cite{Li:1992nu,Efremov:1979qk}, the proton form factors ~\cite{Aznaurian:1979zz,Li:1992ce,Duncan:1979hi,Lepage:1979za} and exclusive processes such as $B$ decays~\cite{Lepage:1979zb,Lepage:1980fj,Li:1994iu,Li:2012nk}. A good introduction to hadronic form factors in perturbative QCD is given in~\cite{Sterman:1997sx}.

In this paper, we will show how the generic rapidity-renormalized LFWFs can be
obtained from LaMET in a way similar to the transverse-momentum dependent parton distribution functions (TMDPDFs)~\cite{Ji:2019sxk}. The organization of the paper is as follows.  In Sec.II, we review the LF quantization and its conceptual difficulties, especially the rapidity divergences. We emphasise that there is an implicit infinite rapidity limit behind the LF quantization which adds additional transcendentality to LF formulation of QFT in comparison to the equal time formulation.  Due to the effective theory nature of LFQ, the rapidity divergences appear in a way similar to the emergence of UV divergences in the continuum limit and require a proper treatment. In the LF formulation, however, it is very difficult to regulate the rapidity divergence consistently due to breaking of Lorentz invariance. We argue that it is simpler to stay in the instant-quantized theory and treat the LF quantities as gauge-invariant correlation functions with rapidity regulators. In the spirit of LaMET, using rapidities of external states as physical off-light-cone regulator, one can obtain the LF quantities without LF quantization.

In Sec.III, we formulate LFWF amplitudes as gauge-invariant correlation functions with light-like gauge-links attached to maintain gauge invariance. These gauge-links lead to rapidity divergences which must be regularized using rapidity regulators. The naive LFWF amplitudes depend on these regulators and contain additional soft contributions. We introduce the generalized soft functions to remove these additional soft contributions and define the physical LFWF amplitudes that can be used in factorization theorems.

In Sec.IV, we present the LaMET formulation of LFWF amplitudes. We first introduce the quasi-LFWF amplitudes defined similar to the quasi parton distribution functions (PDF) and quasi-TMDPDF. The large hadron momentum $P^z$ plays the role of a physical off-light-cone regulator.  At  large $P^z$,  the quasi-LFWF amplitudes can be matched to the physical LFWF amplitudes with  the help  of reduced  soft  functions.  We introduce the generalized off-light-cone soft function and present the factorization theorem of quasi-WF amplitudes. A  sketch  of  its  proof is included.   As  an  application of  the  factorization  formula,  we  show  that  the  Collins-Soper (CS) rapidity evolution kernels  can  be  extracted  from  ratio  of  quasi-WF amplitudes in which the soft function contribution cancels.

In Sec. V, we use the leading LFWF amplitude of a pseudo-scalar meson as an example to demonstrate to illustrate the large momentum expansion formalism. And we conclude the paper in Sec. VI.

\section{Partons and Light-Front Quantization as Effective Theory}

Partons are an idealized concept which has been motivated from high-energy scattering,
in which the constituents of hadrons all travel collinearly at large momenta,
which can be taken to the limit of infinity on the scale of the strong
interaction scale $\Lambda_{\rm QCD}$. These are specialized modes of QCD
whose dynamics can be described by an effective theory of soft and collinear
degrees of freedom from which the individual hadrons can be constructed.
By choosing a particular direction of collinear modes, one can
construct a Hamiltonian formulation of the theory, which has been
called light-front quantization. In this view, LFQ of a theory
is actually an effective theory, just like the heavy quark effective theory in which
a particular set of modes is selected to describe the quark of
infinite mass. As such, the LFQ cannot describe well the soft-gluon physics at very small $k^+$ which are
part of the hadrons.

In this section, we will review the standard formalism
of the light-front quantization and explain why one shall take the
view that it is an effective theory of QCD and hadron structure.

\subsection{Basics of Light-Front Quantization}

As early as 1949, Dirac had advocated three forms of dynamics,
with light-front being one of them~\cite{Dirac:1949cp}.
In light-front theory, one defines two coordinates,
\begin{equation}
      \xi^\pm = (\xi^0\pm \xi^3)/\sqrt{2} \ ,
\end{equation}
where $\xi^+$ is the light-front ``time'', and $\xi^-$ is the
light-front ``spatial coordinate''. And any four-vector $A^\mu$ will be now written
as $(A^+,A^-,\vec{A}_\perp)$.
Dynamical degrees of freedom are defined on the $\xi^+=0$ plane with
arbitrary $\xi^-$ and $\vec{\xi}_\perp$, with conjugate momentum $k^+$
and $\vec{k}_\perp$. Dynamics is generated
by light-cone Hamiltonian $H_{\rm LC}=P^-$. For a free particle,
with 3-momentum $(k^+, \vec{k}_\perp)$, the on-shell LF
energy is $k^-=(\vec{k}_\perp^2+m^2)/2k^+$.

For theories like QCD, the dynamical degrees of freedoms
are defined as $\psi_+$ and $A_\perp$. Defining Dirac matrices
$\gamma^\pm = 1/\sqrt{2}(\gamma^0\pm\gamma^3)$, the
projection operators for Dirac fields are defined
as $P_\pm = (1/2)\gamma^\mp\gamma^\pm$. Any Dirac field $\psi$
can be decomposed into $\psi=\psi_++\psi_-$ with
$\psi_\pm = P_\pm \psi$, and $\psi_+$ is considered as a dynamical degree of freedom. For the gauge field, $A^+$ is fixed by choosing the LF gauge
$A^+=0$ and $\vec{A}_\perp$ are dynamical degrees of freedom. The physics of the LF correlation becomes manifest if one introduces the LF quantization conditions for dynamaical fields
\begin{eqnarray}
& \left\{\psi_+^\dagger(\vec{\xi}),\psi_+(0)\right\} = P_+\delta^3(\vec{\xi}) \ , \\
& \left[A^i(\vec{\xi}), \partial^+A^j(0)\right]=\frac{i}{2}\delta^{ij}\delta^3(\vec{\xi}) ,
\end{eqnarray}
where three-vectors and delta-functions are all in the sense of LF
coordinates. To solve the commutator relation, one starts with the canonical expansion,
\begin{align}\label{eq:LFquark}
  & \psi_+(\xi^+=0,\xi^-,\xi_\perp)
  = \int \frac{d^2k_\perp}{ (2\pi)^3}
    \frac{dk^+}{ 2k^+}\sum_\sigma \Big[ b_\sigma(k) u(k\sigma)
     \nonumber \\
 &  \times e^{-i(k^+\xi^--\vec{k}_\perp\cdot\vec{\xi}_\perp)}
  + d_\sigma^\dagger(k) v(k\sigma)e^{i(k^+\xi^--\vec{k}_\perp\cdot\vec{\xi}_\perp)}
 \Big]
  \ ,
\end{align}
where $b^\dagger(b)$ and $d^\dagger(d)$ are quark and antiquark creation (annihilation)
operators, respectively. We adopt covariant normalization for the particle
states and the creation and annihilation operators, i.e.,
\begin{eqnarray}
& \big\{b_\sigma(k),b_{\sigma'}^\dagger(k')\big\}=
\big\{d_\sigma(k),d_{\sigma'}^\dagger(k')\big\} \nonumber \\
& =(2\pi)^3\delta_{\sigma\sigma'}2k^+\delta(k^+-k^{'+})
\delta^{(2)}(\vec{k}_\perp-\vec{k}_\perp^\prime)\ ,
\end{eqnarray}
where $\sigma$ is the light-cone
helicity of the quarks which can take $+1/2$ or $-1/2$. We ignore
the masses of the light up and down quarks.

Likewise, for the gluon fields in the light-cone gauge $A^+=0$,
$\vec{A}_\perp$ is dynamical and has the expansion,
\begin{align}\label{eq:LFgluon}
  &  \vec{A}_\perp(\xi^+=0,\xi^-,\xi_\perp)
  =  \int \frac{d^2k_\perp}{(2\pi)^3}
    \frac{dk^+}{ 2k^+}
    \nonumber \\
 & \times \sum_\lambda  \Big[ a_\lambda(k) \vec{\epsilon}_\lambda(k)e^{-i(k^+\xi^--\vec{k}_\perp\cdot\vec{\xi}_\perp)}
  + {\rm h.c.} \Big] \ .
\end{align}
And we have the following covariant normalization for the creation and annihilation operators for gluon,
\begin{equation}
\left[a_\lambda(k),a_{\lambda'}^\dagger(k')\right]=(2\pi)^3\delta_{\lambda\lambda'}2k^+\delta(k^+-k^{'+})
\delta^{(2)}(\vec{k}_\perp-\vec{k}_\perp^\prime) \ .
\end{equation}
Implicitly, the gauge fields $A^\mu$ is a traceless $3\times 3$
matrix with $A^\mu=\sum_aA^{a\mu}T^a$, where $T^a$ are the
$SU(3)$ Gell-Mann matrices satisfying $[T^a,T^b]=if^{abc}T^c$
and $\{T^a,T^b\}=\frac{1}{3}\delta_{ab}+d_{abc}T^c$, where $f^{abc}$ and $d^{abc}$ are the group constants.

Given $\psi_+$ and $\vec{A}_\perp$, using equation of motions $\psi_-$ and $A^-$ can be expressed in terms of $\psi_+$ and $A_\perp$ ~\cite{Kogut:1969xa}.  In terms of these light-cone free fields, the LF Hamiltonian can be decomposed into a free part $H_{\rm LC}^{\rm free}$ and an interaction part $V+V_{\rm instan} $
\begin{align}
    H_{\rm LC}&=\int d\xi^-d^2\vec{\xi}_\perp T^{-+}(\xi^+=0,\xi^-,\vec{\xi}_\perp) \ , \nonumber \\
    &=H_{\rm LC}^{\rm free}+V+V_{\rm inst} \ ,
\end{align}
where $H_{\rm LC}^{\rm free}$
is the free kinematic energy on the light-cone, and the interaction $V$ contains the standard quark-gluon vertex and the $3$-gluon, $4$-gluon interactions. The new feature of the light-cone quantization lies in the {\it instantaneous interactions} $V_{\rm inst}$~\cite{Brodsky:1997de} similar to the static Coulomb interactions in Coulomb gauge quantum electrodynamics.

Given the LF Hamiltonian, bound states in QCD can be formulated in a way similar to standard eigenvalue problem in quantum mechanics. The expansion coefficients of the hadroinc functions in the above free Fock states are called LFWF amplitudes. Perturbatively, one can apply the old-fashioned perturbation theory to $H_{\rm LC}$ to calculated the LFWFs. The resulting perturbative series is called light-front perturbation theory (LFPT) and can be formally obtained from Feynman perturbation theory in light-cone gauge by integrating out $k^-$ first~\cite{Collins:2011ca}. A major feature of LFPT is that naively looking, due to the $k^+\ge 0$ constraints, no particle can be created out of or annihilated into the vacuum, therefore there is a clear separation between particles with $k^+>0$ and the vacuum. This is in sharp contrast with the equal-time quantization where particles can be created from vacuum and there is no clear separation between particles and the vacuum.

The hope of LFQ for QCD is not about perturbation theory, but to solve the hadron states on the light-front~\cite{Brodsky:1997de}.
The discretized light-cone quantization was proposed in~\cite{Pauli:1985ps} to solve the bound state problem.
This non-perturbative framework allows one to treat the zero-mode problem explicitly in~\cite{Heinzl:1991myy,Burkardt:1992sz,Yamawaki:1998cy} and turns out to be successful for models in 1+1 dimension, such as the Schwinger
model~\cite{Heinzl:1991myy,McCartor:1994im,Harada:1995au,Kalloniatis:1996db}, the 1+1 QCD~\cite{Zhitnitsky:1985um,Burkardt:1989wy,Srivastava:2000cf},
the 1+1 $\phi^4$ theory~\cite{Harindranath:1987db} and the sine-Gordon model~\cite{Burkardt:1992sz}.

\subsection{Zero Modes, Rapidity Divergences and LFQ as Effective Field Theory}

However, as one realized later, the simplification due to LFQ is not as trivial as one might thought about at a first glance.  Even in $1+1$ dimensions, the triviality of the vacuum is invalid~\cite{Nakanishi:1976vf} due to the presence of non-negligible light-cone zero modes (modes with $k^+=0$). They corresponds to long-wave length fluctuations at large light-cone separation and are sensitive to the vacuum structure~\cite{Yamawaki:1998cy}. A proper treatment of such modes requires an IR regulator such as a finite box in $\xi^{-}$ direction that was adopted in the ``discrete light-cone quantization''~\cite{Pauli:1985ps}.  In LFPT, naively neglecting light-cone zero modes will also leads to inconsistent results for certain Feynman diagrams where the $+$ components of certain external momenta become zero. For example, as argued in Ref.~\cite{Collins:2018aqt}, the vacuum bubble diagram for the 1+1 $\phi^4$ theory is non-zero at $P^+=0$. This contribution is entirely due to light-cone zero modes. Integrating out the $k^-$ at $P^+=0$ in a way that leads to LFPT will omit the zero modes contribution and produce incorrect result.

The light-cone zero modes problem that serves as a conceptual ``back-door'' against vacuum triviality of LFQ is not the only severe problem of LFQ. It has been found~\cite{Wilson:1994fk} that the standard power-counting method that works in equal-time or Euclidean formalism failed to produce a simple pattern of UV divergences in LFQ.
In fact, in 4-D gauge theory, a new type of divergence at small $k^+$ called ``light-cone divergence'' appears due to presence of $\frac{1}{k^+}$ in the phase space measures and in the instantaneous vertices. In LFPT, the individual diagrams can diverge even more than logarithmically.  One might think that the light-cone divergence is simply an artifact of LFQ and should cancel at the final step of calculation for physical quantities such as $S$ matrix elements. However, there are quantities for which the cancellation of light-come divergences are not complete. Among them are LFWF amplitudes
and the associated eigenvalue Eq.~(\ref{eq:LFeigen}). We will show in Sec.~\ref{Sec:LFWF} that by expressing the LFWF amplitudes as gauge-invariant correlations functions in covariant gauge, one can identify the non-cancelling light-cone divergences as the famous {\it rapidity divergences}  known in the literature of transverse momentum dependent (TMD) phenomenon~\cite{Collins:2011zzd,Vladimirov:2016qkd,Vladimirov:2017ksc}. In covariant gauge, they are caused by light-like gauge links extending to infinities and can be regulated efficiently by introducing {\it rapidity regulators} to the gauge-links, but in LFQ and LFPT they appears in all diagrams.
The appearance of the light-cone or light-front divergences is a signal that LFQ theory is an effective one
in the sense that the theory serves to emphasizes the infinite-momentum collinear modes.

Due to the above reasons, LFPT has not been used
for any beyond one-loop calculations, except for two-loop anomalous magnetic
moment in QED~\cite{Langnau:1992zj}. In fact, the common wisdom of using dimensional regularization (DR) for the transverse-momentum integral, and cut-off for longitudinal momentum has not been proven useful for multi-loop calculations. However, a successful use of the LFPT has been the derivation of the Balitsky-Fadin-Kuraev-Lipatov evolution~\cite{Kuraev:1976ge,Balitsky:1978ic,Lipatov:1985uk} by Mueller for quarkonium wave functions, for which the rapidity divergence structure is relatively simple~\cite{Mueller:1993rr}.

Due to the complication caused by the LF divergences,  there appear infinite number of counter terms in the LF Hamiltoinian~\cite{Wilson:1994fk},
which describe the interactions of zero-modes with non-zero modes. This is the price that one pays for an effective theory to truncate away the zero modes which contain infrared physics as well. Even without these problems,
one has to use a severe truncation in the number of Fock states to solve  Eq.~(\ref{eq:LFeigen}).
Such truncation usually breaks simple rotational symmetry in the sense that the  states belonging the same angular momentum multiplets will have different energies. Moreover, there has been no demonstration so far that the Fock truncation converges well in QCD~\cite{Wilson:1994fk}.

Thus, the LFQ provides a great way
to understand the parton physics, but is difficult to solve
it directly due to the complicated LF divergences and their renormalization.

\subsection{An Effective Approach to Light-Front Quantization}

We have emphasized that LFQ is an effective theory of high energy scattering in
which the infinite-momentum limit is taken before UV renormalization. This is also the spirit of the soft-collinear effective theory (SCET)~\cite{Bauer:2000yr,Bauer:2001yt}. On the other hand,
one can perform UV renormalization first before taking
the infinite-momentum limit. One can obtain the former
result from the latter by simple EFT matching.
This is the spirit of the large momentum expansion
or effective theory~\cite{Ji:2013dva,Ji:2014hxa}.

More specifically, there is a procedure to obtain the LF correlators from Euclidean correlation functions in instant formalism.  Let's consider for simplicity the following two-point function inside a fast-moving haron state $|P\rangle$ with large but finite hadron momentum $P^{\mu}$
\begin{align}
 \widetilde f_{O} (\lambda,\vec{b}_\perp, \zeta_\xi=\frac{\xi \cdot P }{\sqrt{\xi^2}},\mu)=\langle P|O(\xi)O(0)|P \rangle  \ ,
\end{align}
where $\lambda=\xi\cdot P$ is the longitudinal invariant length and $\vec{b}_\perp$ is the transverse separation. The large rapidity gap between the separation $\xi$ and the hadron is characterized by the variable $\zeta_\xi=\frac{\xi \cdot P }{\sqrt{\xi^2}}$ that plays the role of a hard scale as well. In order to obtain the corresponding LF version $f_{O}(\lambda, \vec{b}_\perp,\zeta,\mu)$ at rapidity scale $\zeta$, two non-trivial operations need to be performed to obtain the corresponding light-cone version \begin{itemize}
\item An operation in UV which removes the contributions in $\widetilde f$ that are due to fluctuations at the hard scale $\zeta_\xi$. This process is usually called {\it matching}. After performing the matching, the hard scale $\zeta_\xi$  dependency ``transmutes'' into the renormalization scale $\mu$ dependence of the LF-correlator.
\item An operation in small $k^+$ region, the {\it rapidity renormalization}, that removes all the contributions form the small $k^+$ fluctuation.  After rapidity renormalization, the physical rapidity dependence ``transmutes'' into the rapidity renormalization scale $\zeta$ dependence of the LF-correlator.
\end{itemize}
Schematically we have the relation :
\begin{align}
&f_O(\lambda,\vec{b}_\perp, \zeta,\mu)\nonumber \\
&=\lim_{\zeta_v \rightarrow \infty} Z_{\rm RD}(\frac{\zeta}{\zeta_\xi})\otimes C_{\rm UV}(\frac{\zeta_\xi}{\mu})\otimes \widetilde f_O(\lambda,\vec{b}_\perp,\zeta_\xi,\mu) \ .
\end{align}
The $Z_{\rm RD}$ is the rapidity renormalization factor and the $C_{\rm UV}$ is the matching kernel. Therefore, in this sense {\it the light-cone theory is obtained from the full theory by ``integrating out'' UV modes above $\zeta_v$ and the small $k^+$ modes below $\Lambda_{QCD}$, and can be viewed as an effective theory to the full theory}.  As expected, the LF quantum fields $\phi$ are not the original quantum fields of the full theory.

\section{Wave-Function Amplitudes and Rapidity Divergences}\label{Sec:LFWF}

In this section, we first express the naive WF amplitudes as LF correlation functions between hadron state and the QCD vacuum, in which light-like gauge-links extending to infinities are required to maintain gauge-invariance.  This also allows the identification of LF divergences as the rapidity divergences, known in the literature of TMD physics, which we will review after introducing the LF amplitudes.  We then introduce the generalized soft functions composed of $N+1$ Wilson-line cusps that remove the regulator dependency in naive LF amplitudes (or defining a universal regulator scheme consistent with
DR). The physical LFWF amplitudes and their evolution equations will also be discussed.

\subsection{WF Amplitudes as LF Correlation Function}

As we have pointed out before, the WF amplitudes can be expressed in the form of Eq.~(\ref{eq:LFcorre}). By inverting the LF creation-annihilation operators in term of LF fields in Eqs.~(\ref{eq:LFquark}) and (\ref{eq:LFgluon}), one can express the WF amplitudes as LF correlation functions with quark and gluon fields separated in $\vec{\xi}_\perp$ and $\xi^-$ directions. Here we present a precise definition of
the LFWF amplitudes in terms of the matrix elements
of non-local LF operators between the hadron states and the QCD vacuum.

We first introduce the dimensionful light-cone vectors $p=\frac{1}{\sqrt{2}P^+}(1,0,0,1)$ and $n=\frac{P^+}{\sqrt{2}}(1,0,0,-1)$ in convention $A^\mu=(A^0,A^1,A^2,A^3)$.
They satisfy $p\cdot n=1$. $P^+$ is a mass dimension-1 parameter which will be chosen appropriately in specific applications. We use a generic notation $\phi_i$ to denote dynamical parts of the quark and gluon fields $\psi$ and $A^\mu$, with indices $i$
to label field types, and any other features such as color representation, flavor, etc.
We introduce gauge-invariant version of the field $\Phi_i$ which contains gauge-link along the light-cone direction $n$, pointing to positive or negative infinity:
\begin{align}
\Phi^{\pm}_{i}(\xi)=W_{n}^{\pm}(\xi)\phi(\xi) \ ,
\end{align}
with light-like gauge-link $W_{n}^{\pm}(\xi)$ defined as
\begin{align}
   W_{n}^{\pm}(\xi)= {\cal P}{\rm exp}\left[-ig\int_{0}^{\pm \infty} d\lambda n\cdot A(\xi+\lambda n)\right] \ ,
\end{align}
where ${\cal P}$ is a path order.
In the LF gauge $n\cdot A=0$, the gauge-links formally disappears and the $\Phi_i$ simply reduces to local LF fields $\phi_i$ in Eqs.~(\ref{eq:LFquark}) and (\ref{eq:LFgluon}). From these fields, one can construct the generic naive LFWF amplitudes,
\begin{align}\label{eq:naiveLFa}
&\psi^{\pm0}_{N}(x_i,\vec{b}_{i\perp},\mu)=\int \prod^N_{i=1} d\lambda_i e^{i\lambda_i x_{i}}\times e^{+i\lambda_0x_0} \\ &\times\langle 0 | {\cal P}_N\prod^N_{i=1}\Phi^{\pm}_{i}(\lambda_{i}n\!+\!\vec{b}_{i\perp})\Phi^{\pm}_0(\lambda_0n\!+\!{\vec b}_{0\perp}) |P\rangle \ . \nonumber
\end{align}
where $x_0=1 - \sum_i x_i$, $\lambda_0=-\sum_i\lambda_i$, and all longitudinal momentum fractions carried by partons satisfy $0<x_i<1$. We single out a field labelled with 0 with a fixed longitudinal coordinate $\lambda_0$ and conjugation momentum $x_0$. We choose
the sum of the all longitudinal coordinate $\lambda_i$ as $0$ due to translational symmetry. Likewise, when $\vec{P}$ has no
transverse momentum $\vec{P}_\perp =0$, the transverse coordinate $\vec{b}_\perp$
can be shifted by an overall constant without any effect.
The 0 superscript on the amplitude indicates that rapidity divergences have not been regularized. However, regular UV divergences
are regularized in dimensional regularization (DR) with renormalization scale $\mu$ and in the modified minimal subtraction ($\overline {\rm MS}$)-scheme.

\begin{figure}
\includegraphics[width=0.6\columnwidth]{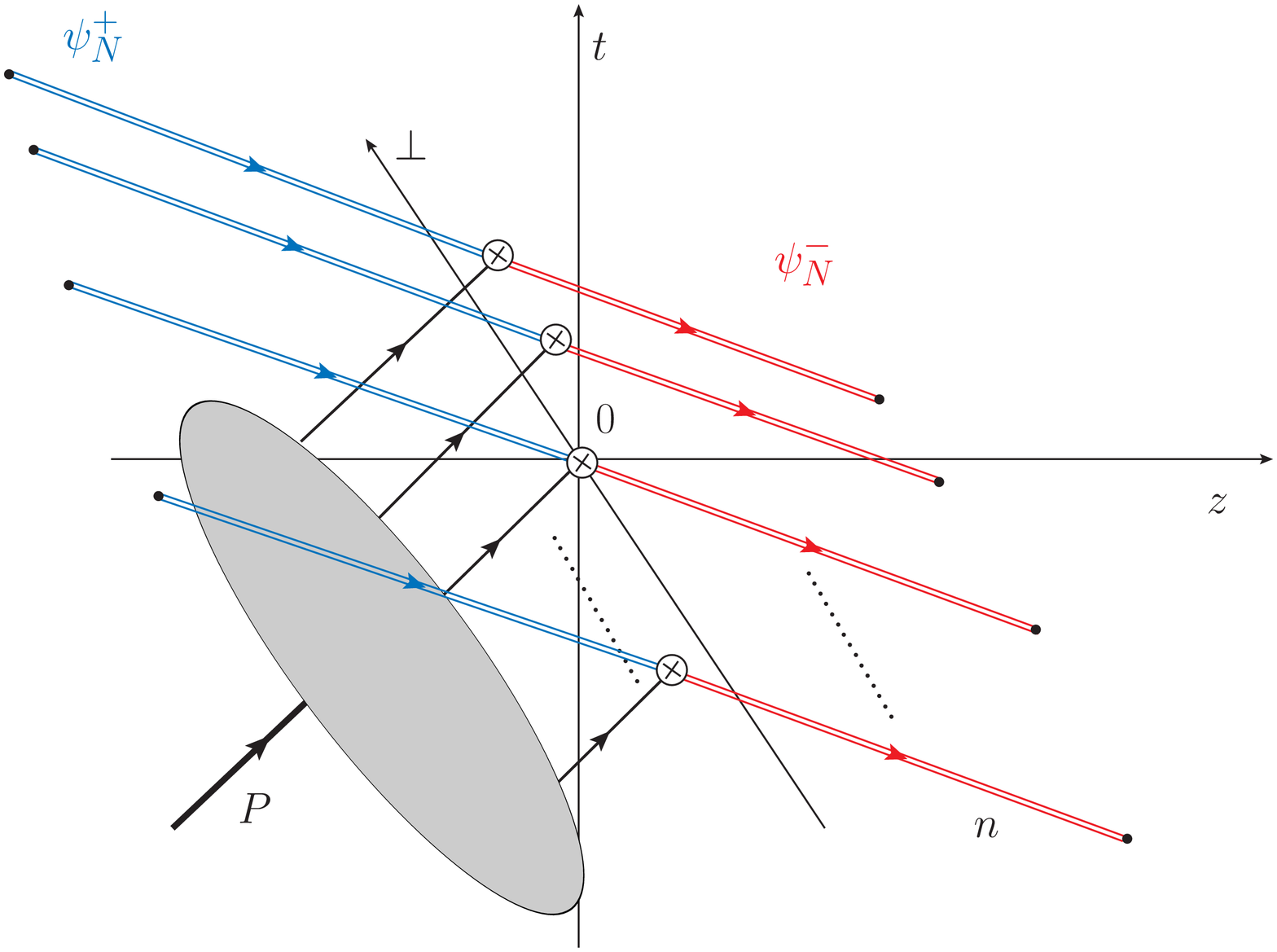}
\caption{\label{fig:WFN} The LFWF amplitudes $\psi_N^{+0}$ (blue) and $\psi_N^{-0}$ (red). The crossed circles denotes the operator insertion $\Phi_i$ or junctions of gauge-links. The explicit form of the junctions of the gauge-links are light-cone infinities is irrelevant as far as gauge-invariance is preserved.  }
\end{figure}
All fields are properly coupled to the quantum numbers of the hadron under consideration. The projection operator ${\cal P}_N$ is to project all the color indices onto the singlet channel.
For example, for $N=N_c-1$ and all the $\Phi_i$s are in the fundamental representation, the projection is the total anti-symmetrization while for $N=1$ and one fundamental, one conjugate fundamental, the projection is the trace. It must be distinguished from the path-ordering operator ${\cal P}$.
There may be different ways to couple the same set
of fields into the required quantum numbers and they are treated as independent.
We have also omitted the helicity and angular momentum coupling to generate
a specific helicity combination~\cite{Ji:2002xn}. See Fig.~\ref{fig:WFN} for a depiction of these LFWF amplitudes.

The above amplitude is gauge-invariant without the transverse gauge-link at light-cone infinity if calculated in non-singular such as the covariant gauge. However, in light-cone gauge $A^+=0$, the gauge potential
does not vanish at infinity, one must specify connections of
the gauge-links at $\lambda=\pm \infty$~\cite{Belitsky:2002sm}.
A similar observation is true for calculation in a finite spacetime volume with a fixed boundary.
The choice of connection method does not affect the relative amplitude between partons with $k^+\ne 0$, but will affect the overall normalization of the amplitudes through the effects of the zero modes as boundary conditions at infinity can only affect zero modes.

One can further Fourier-transform the above amplitudes
to the transverse momentum space. Since the matching formula
to lattice is much simpler in the cooridinate space, it is usually done at the end of a calculation.

\subsection{Rapidity Divergence and Rapidity Regulators}

Neglecting all the divergences, one can formally show that the naive LF amplitudes in Eq.~(\ref{eq:naiveLFa}) reduces to the ones in the hadron wave function defined in light-cone gauge. The naive amplitudes in Eq.~(\ref{eq:naiveLFa}), however, suffers from a new type of divergence associated with the infinitely-long light-like gauge-links. These divergences are due to radiation of gluons collinear to the light-like gauge-link and cannot be regularized by the standard UV regulators. An example is
the following integral in dimensional regularization (DR)~\cite{Ebert:2019okf},
\begin{equation}
     I = \int dk^+dk^- \frac{f(k^+k^-)}{(k^+k^-)^{1+\epsilon}} = \frac{1}{2}
     \int \frac{dy}{y} \int dm^2 \frac{f(m^2)}{m^{2+2\epsilon}}\ ,
\end{equation}
where $m^2=k^+k^-$ and $y = k^+/k^-$ is the rapidity-related variable. The divergences
in $y$ arise from large and small $y$ where the integral is unregulated.  It is clear that this divergence is precisely the non-cancelling part of the
light-cone divergence at small $k^+$ when Eq.~(\ref{eq:naiveLFa}) was evaluated in LFPT. However, in LFPT such divergences scatter across all the diagrams and can not be regulated consistently. In the manifestly Lorentz-covariant formalism, it is purely caused by the gauge-link and is much easier to keep track with.
This rapidity divergence is the signature that partons are objects in effective field theories.

To regulate the light-cone or rapidity divergences, a number of methods have been introduced in literature (for a review see~\cite{Ebert:2019okf}).
They can be put into two classes: on-light-cone regulators and off-light-cone regulators.
In the former case, the gauge-links are kept along the light-cone direction $n^\mu$ after regularization. For example, the so-called {\it $\delta$ regulator}~\cite{Echevarria:2015usa,Echevarria:2015byo} regularizes the gauge-link as:
 \begin{align}
&W^{\pm}_n(\xi)\rightarrow W^{\pm}_{n}(\xi)|_{\delta^-}\nonumber \\
&={\cal P}{\rm exp}\left[-ig\int_0^{\pm \infty} d\lambda n\cdot A(\xi+\lambda n)e^{-\frac{\delta^-}{2} |\lambda|}\right] \,,
\end{align} and similarly for the conjugate direction, where $\delta^-$ is a {\it positive} dimensionless quantity that regulates the light-cone divergence. The $\delta$ regulator breaks gauge-invariance nominally, but the breaking effects might go to zero smoothly as $\delta^-\rightarrow 0$. The regulator preserves the boost invariance $\delta^{\pm}\rightarrow e^{\pm Y}\delta^{\pm}$ where $Y$ is the rapidity of the (residual) Lorentz boost. The {\it LF-distance regulator}~\cite{Vladimirov:2020umg}, on the other hand, regulates the LF divergence through finite LF distance $L^-$
\begin{align}
&W^{\pm}_n(\xi)\rightarrow W^{\pm}_{n}(\xi)|_{L^-}\nonumber \\
&={\cal P}{\rm exp}\left[-ig\int_0^{\pm L^-} d\lambda n\cdot A(\xi+\lambda n) \right] \ ,
\end{align}
which preserves the gauge-invariance at large $L^-$, but is difficult for analytic calculations, in additional to adding transverse gauge links. Other on-light-cone regulators include the exponential regulator~\cite{Li:2016axz}, $\eta$ regulator~\cite{Chiu:2012ir}, analytical regulator~\cite{Becher:2010tm}, etc. Not all of them work in the case of LFWF amplitudes, since some of them are defined with the presence of intermediate state cuts. Nevertheless, the $\delta $ regulator and the LF-distance regulator can still be implemented in the context of LFWF amplitudes.  In the remainder of this section, we will use the $\delta$ regulator as a representative whenever we need an on-light-cone regulator.

The off-light-cone regulator was introduced in~\cite{Collins:1981uk,Ji:2004wu,Ji:2004xq,Collins:2011zzd}, and also used in~\cite{Ji:2004wu}. This type of regulator chooses off-light-cone directions to avoid the rapidity divergence. One can choose, for instance, to deform the gauge-links into the space-like region:
\begin{align}
n\rightarrow n_Y=n-e^{-2Y}\frac{p}{(P^+)^2}\,.
\end{align}
Here $Y$ plays the role of a rapidity regulator, as when $Y \rightarrow \infty$, $n_Y \rightarrow n$. In certain cases one can also deform $n_Y$ into time-like region~\cite{Collins:2004nx}.

The on-light-cone regulators are consistent with the spirit of parton physics, and therefore are useful to define residual-momentum-independent parton densities. The off-light-cone regulators, on the other hand, are in a similar spirit as LaMET, and therefore can be exploited for practical lattice QCD calculations, as we shall see in the next subsection.

To avoid light-cone divergences, from now on we include the rapidity regulator in the definition of the LFWF amplitudes.  The gauge-invariant fields $\Phi_i$ are regularized  as
\begin{align}
\Phi^{\pm}_{i}(\xi;\delta^-)= W^{\pm}_{n}(\xi)|_{\delta^-}\phi(\xi) \ ,
\end{align}
in terms of which the un-subtracted LFWF amplitudes becomes
\begin{align}\label{eq:naivefullam}
&\psi^{\pm0}_{N}(x_i,\vec{b}_{i\perp},\mu,\delta^-)=\int \prod^N_{i=1} d\lambda_i e^{i\lambda_i x_{i}} \times e^{i\lambda_0x_0}\\ &\times\langle 0 | {\cal P}_N\prod^N_{i=1}\Phi^{\pm}_{i}(\lambda_{i}n\!+\!\vec{b}_{i\perp};\delta^-)\Phi^{\pm}_0(\lambda_{0}n\!+\!{\vec b}_{0\perp};\delta^-) |P\rangle \ . \nonumber
\end{align}
The subscript $\delta^-$ denotes that the gauge-links in $\Phi_i$ are regulated by the $\delta$ regulator in the light-cone minus direction.  Similar to the case of TMDPDFs, as $\delta^- \rightarrow 0$, $\psi$ diverges logarithmically, and the finite part also depends on the rapidity regulator. Therefore, the naive amplitudes in Eq.~(\ref{eq:naivefullam}) can not appear in factorization formulas for physical observable by themselves. One must remove all divergences and rapidity regularization scheme dependencies in $\psi$, in a way similar to removing UV divergences in physical quantities. These can be accomplished with the help of {\it soft functions} to be introduced in the next subsection.

We should mention that the rapidity divergence in gauge theory is not simply an artifact, it reflects the deep fact that as one boost a hadron to faster and faster speed, soft gluons at long-wave length are continually generated with larger and larger population. Therefore, the rapidity divergence is closely related to the small-$x$ physics.  Here we simply mention a famous example of Mueller for the onium wave function. Assuming the initial wave function is given by $\psi_0(x,b_T)$, let us consider the radioactive correction to the wave function at order $\alpha_s^2$. This calculation have been performed by Mueller in \cite{Mueller:1993rr}, see also \cite{Kovchegov:2012mbw}. For the valence component consists of one quark and one anti-quark we simply need the virtual part of his result. The divergent piece is given by:
\begin{align}
&\psi_{\bar q q}(x,x_{\rm min},b_\perp,\mu)|_{\rm div} \nonumber \\
&\approx -\frac{\alpha_sC_F}{\pi}\ln \mu^2b_\perp^2 \int^{1-x}_{x_{\rm min}} \frac{dx_g}{x_g}\psi_0(x,b_T) \ .
\end{align}
The coefficient of the divergence is nothing but the one-loop Collins-Soper evolution kernel for TMDPDF. If we probe reals emissions with Glauber exchanges as normally happens in the real collision, then the rapidity divergence simply can not be captured by a linear evolution equation. Instead, at large $N_c$ the rapidity evolution for the generating functional for the soft-gluon wave functions is controlled by the non-linear Balistky-Kovchegov (BK) equation~\cite{Balitsky:1995ub,Kovchegov:1999ua}, which can be mapped to the Banfi-Marchesini-Smye (BMS) equation~\cite{Banfi:2002hw} associated to those ``non-global'' infrared logarithms. A throughout discussion on the small-$x$ evolution equation and the BMS equation is beyond the scope of this paper.

\subsection{Generalized Soft functions for LFWF Amplitudes}

Similar to the case of TMDPDFs, the rapidity divergences as well as scheme dependencies in Eq.~(\ref{eq:naivefullam}) can be renormalized with the help of generalized soft functions, introducing a new rapidity scale $\zeta$. Specifically, the generalized soft function for the $\psi_N$ in Eq.~(\ref{eq:naivefullam}) is composed of $N+1$ Wilson-line cusp operators in the representation set ${\cal R}=\{R_i; i\in (0,N)\}$ where $R_i$ denote the color representation of the $i$th cusped Wilson-line. The Wilson-line cusp operator is defined as
\begin{align}
&{\cal C}^{\pm}(\vec{b}_\perp,\delta^+,\delta^-)=W^{\pm}_n(\vec{b}_\perp)|_{\delta^-}W^{\dagger}_p(\vec{b}_\perp)|_{\delta^+} \ ,
\end{align}
where $W_p$ is defined as
\begin{align}
&W^{\pm}_p(\vec{b}_\perp)={\cal P}{\rm exp}\left[-ig\int_{0}^{\pm\infty} d\lambda' p \cdot A(\lambda' p+\vec{b}_\perp)\right]  \ .
\end{align}
Here the $\lambda'$ has mass dimension $-2$. And the $\pm$ for the minus direction in $W_n^\pm$ should be chosen the same as that of the WF amplitudes.

\begin{figure}
\includegraphics[width=0.6\columnwidth]{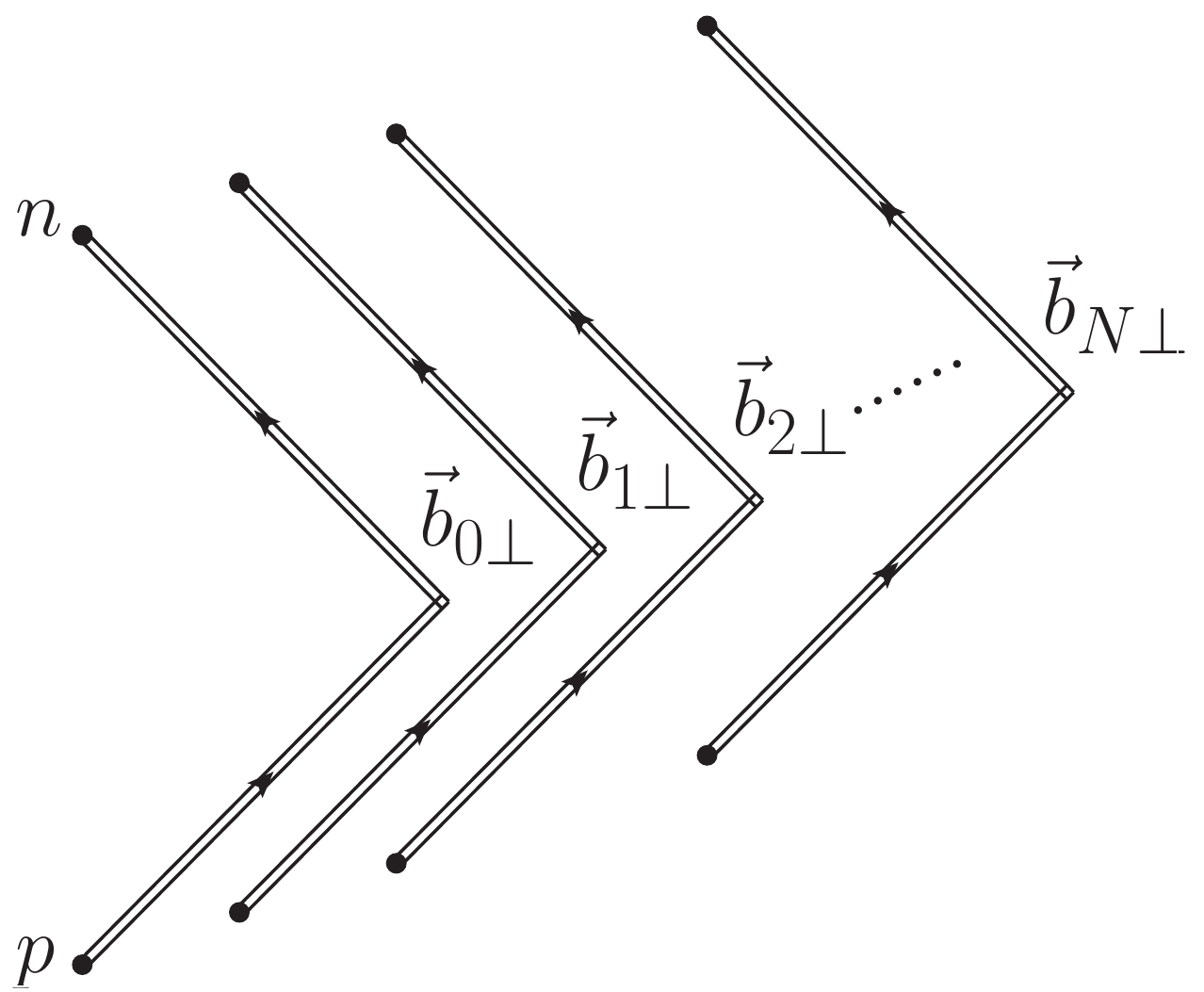}
\includegraphics[width=0.7\columnwidth]{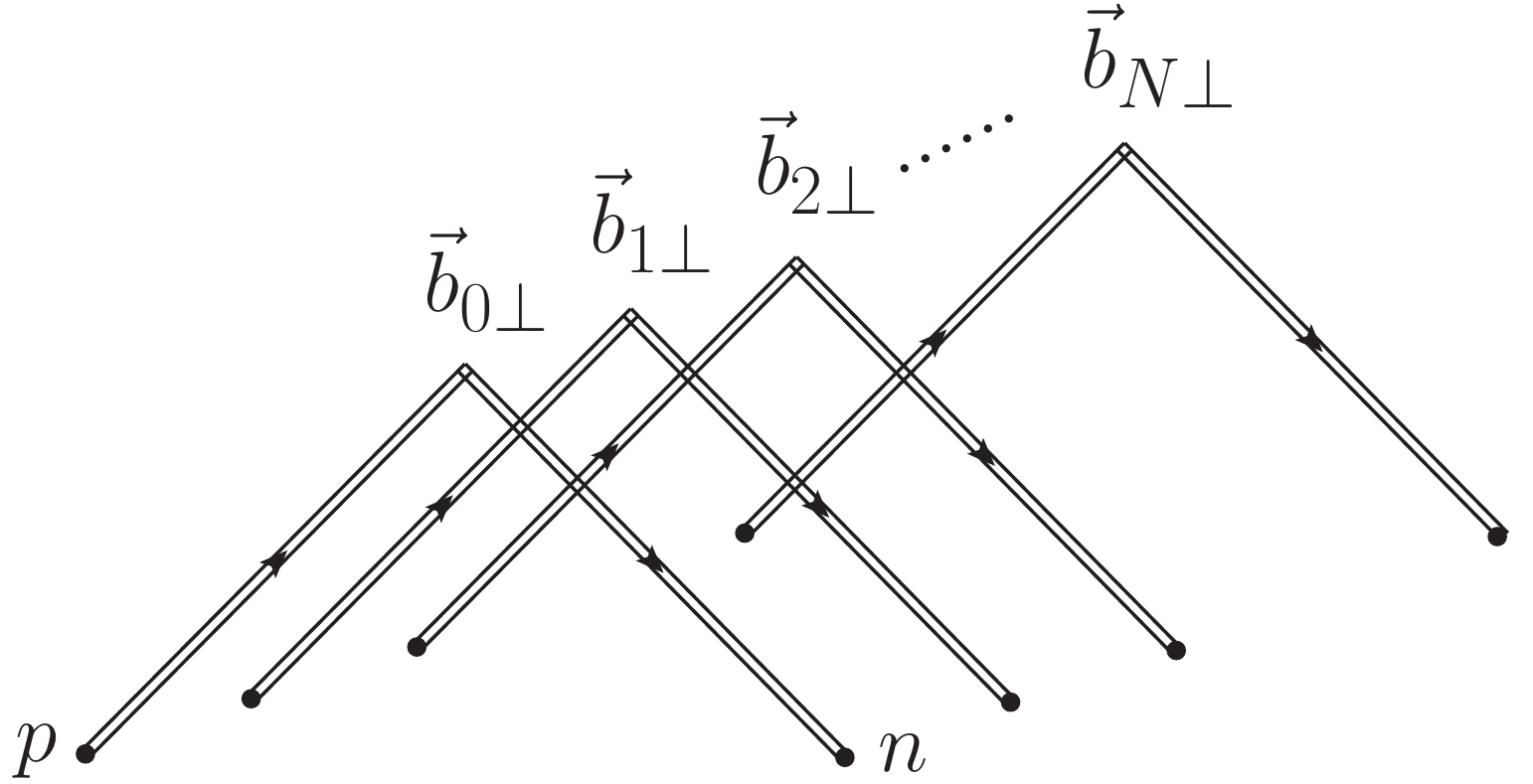}
\caption{\label{fig:SN} The soft function $S_N^+$ (upper) and $S_N^-$ (lower) which can be used
to renormalize the light-cone rapidity divergences in LFWF amplitudes. }
\end{figure}

With the above, we define the generalized soft function as
\begin{align}\label{eq:S_N}
S^{\pm}_{N,{\cal R}}(\vec{b}_{i\perp},\mu,\delta^+,\delta^-)=\langle 0|{\cal T}{\cal P}_N\prod_{i=0}^{N}{\cal C}^{\pm}(\vec{b}_{i\perp},\delta^+,\delta^-)|0\rangle \ ,
\end{align}
where ${\cal T}$ is a time-ordered product for quantum fields. Notice that the time and path orderings are operating on different spaces of operators. For $N=1$ and ${\cal R}=\{f,\bar f \}$ where $f$,$\bar f$ denote fundamental and conjugate fundamental representations, the definition reduces to the standard TMD soft function for quark-TMDPDF~\cite{Collins:2011zzd}. See Fig.~\ref{fig:SN} for a depiction of the above generalized soft function. In the following discussion, we will always omit the label ${\cal R}$ for the color-representation of the generalized soft functions unless otherwise mentioned. We will also omit the ``generalized '' from their names and call the case with $N=1$ and $N>1$ generically as soft functions.

Intuitively, the soft functions are obtained from the WF amplitudes by performing eikonal approximations to the incoming parton lines. They re-sum all the soft-gluon radiations from the bare WF amplitudes and suffer from rapidity divergences. Therefore, the generic rapidity regulator is also imposed on the soft function.  Since the soft function contains two light-like directions, the scheme dependencies of the soft function are expected to ``double'' that of the WF amplitudes, therefore a square root is introduced in the next subsection to ensure the renormalized WF amplitude is scheme independent.

Indeed, it has been argued in~\cite{Vladimirov:2017ksc} based on conformal transformation that the rapidity divergences in the un-subtracted WF amplitude and the soft functions are indeed multiplicative and is controlled by the generalization of the {\rm Collins-Soper kernel}, labelled here by  $K_N(\vec{b}_{i\perp},\mu)$~\cite{Collins:1981uk}. As $\delta^{\pm} \rightarrow 0$, one has
\begin{align}\label{eq:SNon}
    &S^{\pm}_{N}(\vec{b}_{i\perp,},\mu,\delta^+,\delta^-)\nonumber \\
    &=\exp \left[K_N(\vec{b}_{i\perp},\mu)\ln \frac{\mu^2/(P^+)^2}{\mp 2\delta^+\delta^--i0}+{\cal D}_{2,N}(\vec{b}_{i\perp},\mu)\right] \ ,
\end{align}
where only the product $\delta^+\delta-$ appears due to boost invariance and $(P^+)^2$ comes from the definition of dimensionful light-cone vectors. In principle, the mass parameters in the light-cone vectors can be chosen differently from that of the hadron, but here we chose them to be the same for simplicity. As a result of this choice, the hadron momentum dependency is entirely transmuted  to the soft functions. The generalized rapidity evolution kernel depends on the color representations ${\cal R}$ and reduces to the standard Collins-Soper kernel for $N=1$ and ${\cal R}=\{f,\bar f \}$. $K_N(\vec{b}_{i\perp},\mu)$ is independent of rapidity regularization scheme. On the other hand, the finite part ${\cal D}_{2,N}(\vec{b}_{i\perp},\mu)$ (where subscript 2 indicates the definition depending on the two on-light-cone vectors) is scheme-dependent.

Similar to the $N=1$ case,  the soft-function in $\delta$ regularization satisfies the renormalization group equation
\begin{align}
&\mu^2\frac{d}{d\mu^2}\ln S^{\pm}_{N}(\vec{b}_{i\perp,},\mu,\delta^+,\delta^-)\nonumber \\
&=-\frac{N+1}{2}\Gamma_{\rm cusp}(\alpha_s)\ln \frac{\mu^2/(P^+)^2}{\mp 2\delta^+\delta^--i0} + \frac{N+1}{2}\gamma_s(\alpha_s)\ ,
\end{align}
where $\Gamma_{\rm cusp}(\alpha_s) $ is the light-like cusp anomalous dimension~\cite{Polyakov:1980ca,Korchemsky:1987wg} and the $\gamma_s(\alpha_s)$ is the soft anomalous dimension independent of $N$~\cite{Korchemskaya:1992je}. But they all depends on the color representation ${\cal R}$. The generalized Collins-Soper kernel satisfy the renormalization group equation (RGEs):
\begin{align}
\mu^2\frac{d}{d\mu^2} K_N(\vec{b}_{i\perp},\mu)=-\frac{N+1}{2}\Gamma_{\rm cusp}(\alpha_s) \label{eq:cusp} \ .
\end{align}
It is worth pointing out that $S_N^{-}$ and $K_N$ in some special cases have been calculated to two loops in~\cite{Vladimirov:2016qkd}.

One should notice that the $\pm$ sign in the logarithms in Eq.~(\ref{eq:SNon}) is related to the analyticity structure of the soft functions. Notice that the $e^{-\lambda \delta^{\pm}}$ in the definition of the delta-regulator can be viewed as residual imaginary external momenta $\delta k_p=i\delta^{-}p$ and $\delta k_n=i\delta^{+} (P^+)^2n $ carried by the corresponding gauge-links in $p$ and $n$ direction, respectively. For the $S_N^-$ space-time picture, both of them are incoming and the residual momentum transfer equals to $Q^2=(\delta k_p+\delta k_n)^2=-(P^+)^2\delta^+\delta^-$, which is space-like. Therefore the corresponding soft functions $S_N^-$ are purely real.  On the other hand, for the $S_N^+$ picture one residual momentum is incoming while another is out-going and the residual momentum transfer $(\delta k_p-\delta k_n)^2$ becomes time-like, which will generate an imaginary part. In fact, one can show that the two versions of the soft functions are related through analytic continuation
\begin{align}
    S^{+}_N(\vec{b}_{i\perp},\mu,\delta^+\delta^-)=S^{-\dagger}_N(\vec{b}_{i\perp},\mu,-\delta^+\delta^-+i0) \ .
\end{align}
This is similar to the relation between space-like and time-like form-factors~\cite{Collins:2016hqq} where the momentum transfer $Q^2$ equals to $-(P^+)^2\delta^+\delta^-$.

\subsection{Rapidity-Renormalized LFWF Amplitudes and Rapidity-Scale Evolutions}

With the above soft functions, we take its square root to perform the rapidity renormalzation of $\psi^0_N$ in Eq.~(\ref{eq:naivefullam}) and define the ``physical'' LFWF amplitudes as
\begin{align}\label{eq:physicalWF}
&\psi^{\pm}_N(x_i,\vec{b}_{i\perp},\mu,\zeta)=\lim_{\delta^- \rightarrow 0}\frac{\psi^{\pm 0}_N(x_i,\vec{b}_{i\perp},\mu,\delta^-)}{\sqrt{S^{\pm}_{N,{\cal R}}(\vec{b}_{i\perp},\mu,\delta^-e^{2y_n},\delta^-)}} \ .
\end{align}
where $S_{N,{\cal R}}$ is the generalized TMD soft function defined in Eq.~(\ref{eq:S_N}).

Similar to the case of TMDPDF~\cite{Collins:2011zzd}, We also introduced a dimensionless rapidity parameter $y_n$ for the renormalized WF amplitude. The rapidity divergences cancel between the bare or un-subtracted WF amplitudes and the soft function, leading to the explicit dependence of WF amplitudes on a set of rapidity scales $\zeta$ defined as
\begin{equation}
\zeta=2(P^+)^2 e^{2y_n} .
\end{equation}
Sometimes it is also useful to introduce  $N+1$ rapidity scales related to $\zeta$ and $x_i$ through
\begin{align}
    \zeta_i\equiv\zeta x_i^2 \ .
\end{align}
Given $x_i$ and $\zeta$, they are not independent variables, but they are the most natural hard scales around the $N+1$ link-field vertices.
The rapidity evolution equation for the rapidity-renormalized WF amplitudes reads
\begin{align}
2\zeta \frac{d}{d\zeta} \ln \psi^{\pm}_N(x_i,\vec{b}_{i\perp},\mu,\zeta)=K_N(\vec{b}_{i\perp},\mu) \ .
\label{eq:kn}
\end{align}
where the generalized Collins-Soper kernel is non-perturbative
for large $\vec{b}_{i\perp}$. The proof of the above
equation is similar to TMDPDF and will not be discussed here. See Refs.~\cite{Collins:1981uk,Ji:2014hxa} for further details.

The RGE for the UV-renormalized WF amplitude can be derived easily,
\begin{align}
&\mu^2\frac{d}{d\mu^2}\ln \psi^{\pm}_N(x_i,\vec{b}_{i\perp},\mu,\zeta)\nonumber \\
&=\frac{1}{4}\sum_{i=0}^{N}\Gamma_{\rm cusp}\ln \frac{\mu^2}{\pm \zeta x_i^2-i0}-\frac{N+1}{2}\gamma^{H}(\alpha_s)  \ ,
\end{align}
where $\gamma_H$ is the hard-anomalous dimension that depends on the specific field operators $\Phi_i$. We will give an example for meson amplitude later.

One should notice that for unpolarized TMDPDFs, the imaginary parts cancel between  different diagrams as they
are real quantities, but for WF amplitudes there are no such cancellations. The imaginary parts are caused by rapidity logarithms of $P^+$ and various rapidity regulators. The proper $\pm$ in the $\pm \zeta_i-i0$ term is determined by the $i0$ prescriptions in the gauge-link propagators for the unsubtracted or bare WF amplitudes and the soft functions. A correct prescription must guarantee its exponential decay at light-front infinities. The standard set of the LFWF amplitudes defined above
can serve in factorization of experimental processes or comparing with various theoretical calculations.

\section{Quasi LFWF Amplitudes and Factorization in LaMET}
\label{sec:LFWF-LFA_LaMET}

The standard LFWF amplitudes are natural quantities to solve
in LFQ if a viable approach can be found to implement a non-perturbative
solution of Eq. (1). The rapidity dependence will appear naturally when the LF Hamiltonian
is implemented with a rapidity regulator, such as finite length in the LF coordinate, without breaking the manifest Lorentz symmetry.
Alternatively, they can be calculated using
the effective large-momentum expansion approach once the instant-form solutions of some Euclidean matrix
elements in a large momentum state are found~\cite{Ji:2020ect}. The goal of this section is to show how to implement this.

We first need to find a Euclidean version of the WF amplitudes which contain
the same collinear and soft physics as that of the LFWF amplitudes.
Similar to the case of TMDPDFs, the collinear part can be taken into account by
boosting the gauge links in the standard amplitudes without time dependence,
and the light-front time dependence can be taken into account by the large rapidity external hadron states. Thus all the ingredients required to reproduce the correct
collinear and soft physics for WF amplitudes are available on a Euclidean lattice.

In this section, we first define the quasi-WF amplitudes in general. We then introduce the reduced soft functions as the rapidity independent part of the generalized off-light-cone soft functions. They are required to cancel the off-light-cone scheme dependencies from the quasi-WF amplitudes and match to the standard LFWF amplitudes. We then discuss the factorization of quasi-WF amplitudes and present the matching formula.

\subsection{Quasi-WF Amplitudes}

\begin{figure}
\includegraphics[width=0.6\columnwidth]{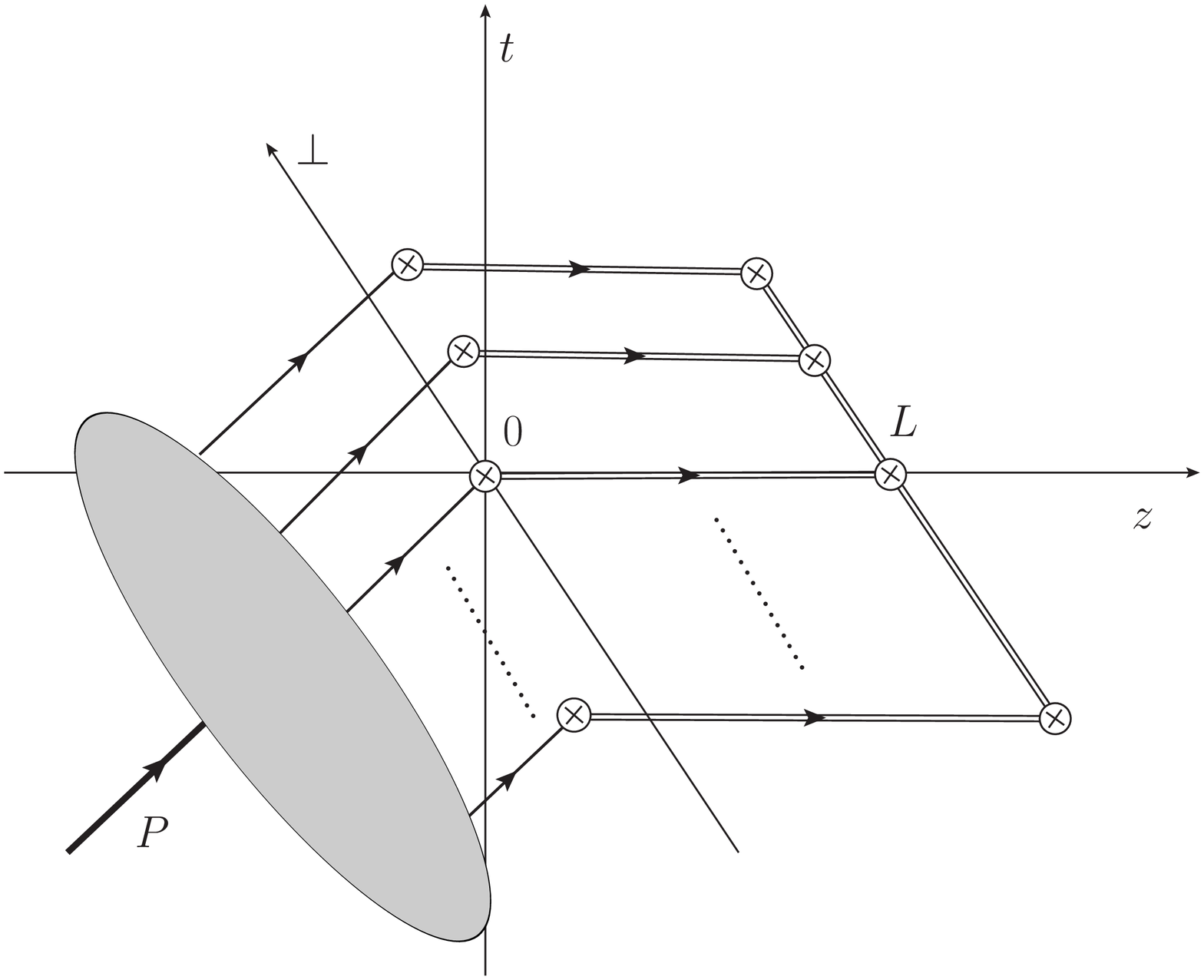}
\includegraphics[width=0.4\columnwidth]{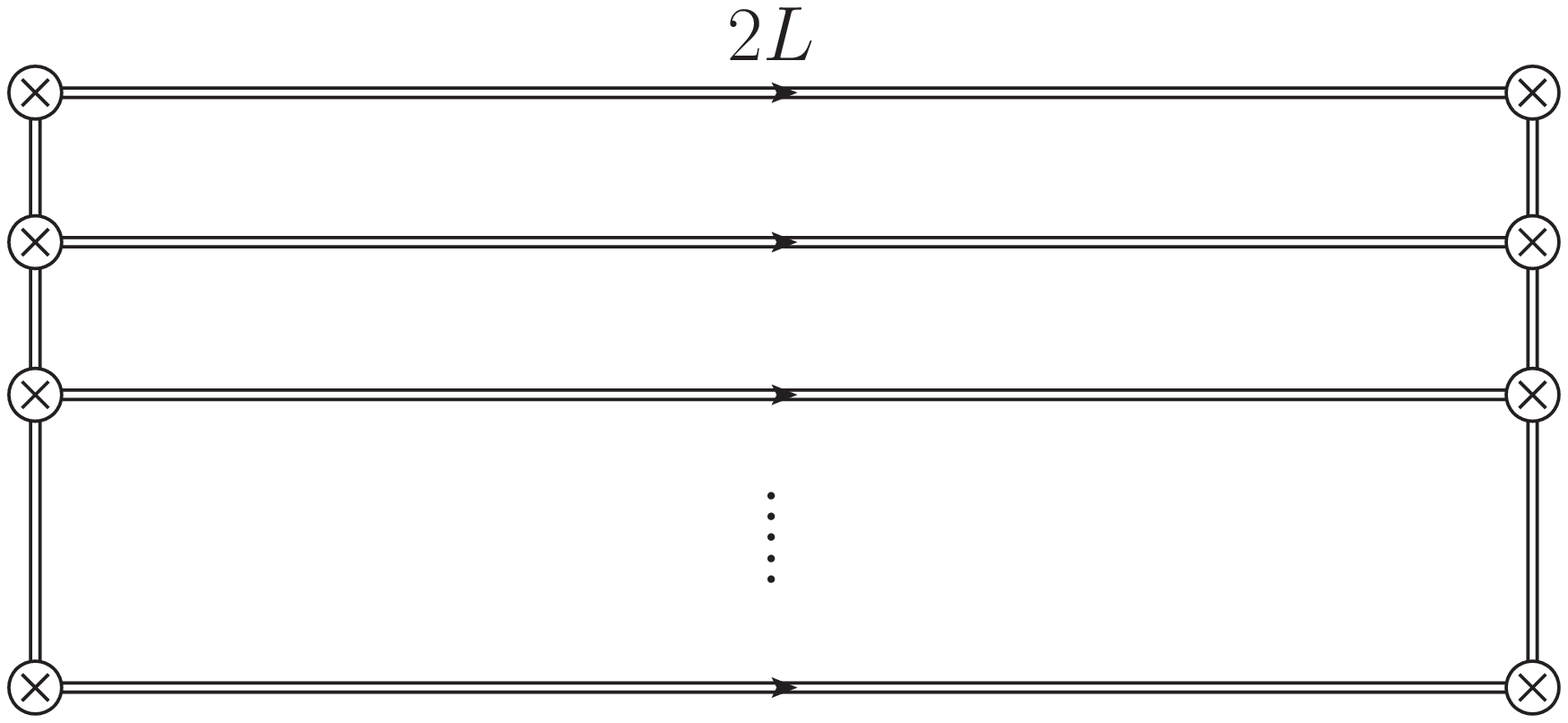}
\caption{ \label{fig:quasi} The quasi-LFWF amplitudes $\widetilde \psi_N^-$ (upper) and $Z_E$ (lower). The crossed circles denotes the operator insertion $\Phi_i$ or junctions of gauge-links. Notice that the connection of the gauge-links at $z=L$ in not unique, however, the contribution at $z=L$ should cancel after taking ratio with $\sqrt{Z_E}$.  }
\end{figure}

Let us denote the unit four-vector in $z$ direction as $n_z=(0,0,0,1)$. We consider the ordinary equal-time (also named as Euclidean) quasi-LFWF amplitudes or simply quasi-WF amplitudes in a large momentum hadron,
\begin{align}\label{eq:quasi_WF}
&\widetilde \psi^{\pm}_N(x_i,\vec{b}_{i\perp},\mu,\zeta_{z})=\lim_{L\rightarrow \infty}\int d\lambda_i e^{-i\lambda_ix_{i}-i\lambda_0x_0} \\ &\frac{ \langle 0| {\cal P}_N\prod^N_{i=1}\Phi^{\pm}_i(\lambda_{i}{n_z}+\vec{b}_{i\perp};L)\Phi^{\pm}_0(\lambda_0 {n_z};L) |P\rangle}{\sqrt{Z_E(2L,\vec{b}_{i\perp},\mu)}} \nonumber
\end{align}
where $\lambda_0=-\sum_{i=1}^{N}\lambda_i$ and  $x_0=1-\sum_{i=1}^{N}x_i$.  The $\Phi^{\pm}_i(\lambda_{i}{n_z}+\vec{b}_{i\perp};L)$ is the gauge-invariant field with gauge-links along $z$ directions (extended to length $L$) being attached
\begin{align} \label{eq:invariantnz}
\Phi^{\pm}_i(\xi;L)={\cal P}{\rm exp}\left[ig\int_{0}^{\mp L\pm \xi^z} d\lambda A^z(\xi\!+\!\lambda {n_z})\right]\phi(\xi) \ ,
\end{align}
with $\xi^z=-\xi\cdot n_z$ and the $\pm L$ corresponds to the $\mp$ choices for the WF amplitude.
The $\zeta_{z}=(2P\cdot {n_z})^2$ are the Collins-Soper rapidity scale similar to that of the quasi-TMDPDFs. One also needs the $N+1$ rapidity scales $\zeta_{z,i}\equiv \zeta_zx_i^2$ similar to the $\zeta_i$.
Clearly the above quantity is the external momentum $P^\mu=(P^0, 0, 0, P^z)$-dependent. The choices of the fields and couplings in the quasi WF are not unique, for a given LFWF amplitude to be reproduced. This is the universality principle
of LaMET discussed in Ref. \cite{Ji:2020ect}. $Z_E(L,b_{i\perp},\mu)$ is the vacuum expectation of a set of space-like Wilson-lines
along $z$ direction and separated in the transverse plane :
\begin{align}\label{eq:ZE}
&Z_E(L,\vec{b}_{i\perp},\mu) \\
&=\langle 0|{\cal P}_N{\cal T}\prod_{i=0}^{N}{\cal P}{\rm exp}\left[ig\int_{0}^{L} d\lambda A^z(\vec{b}_{i\perp}+ \lambda{n_z})\right]|0\rangle \ . \nonumber
\end{align}
The connection in the transverse plane is needed for gauge invariance and might not be unique and shall be in accordance with that
in the standard LFWF amplitudes to be reproduced. The purpose of the factor $Z_E$ in the quasi-LFWF amplitudes
is the same as for quasi-TMDPDFs, in additional to cancel the Wilson line self-energy. See Fig.~\ref{fig:quasi} for a depiction of the quasi-LFWF amplitudes and the $Z_E$.

The external momentum-dependence of the quasi-WF amplitudes can be calculated when the hadron momentum is large. The momentum RG equation can be shown in a way similar to ~\cite{Collins:1981uk} as
\begin{align}
&P^z\frac{d}{dP^z}\ln \widetilde \psi^{\pm}_N(x_i,\vec{b}_{i\perp},\mu,\zeta_{z})\nonumber \\
&=K_N(\vec{b}_{i\perp},\mu)+\sum_{i=0}^N \frac{1}{2}{\cal G}^{\pm }(\zeta_{z}x_i^2,\mu) \ ,
\end{align}
where we have omitted terms of higher powers in $(1/P^z)^2$, and the $K_N(\vec{b}_{i\perp},\mu)$ is the non-perturbative rapidity evolution factor same in Eq. (\ref{eq:kn}) and ${\cal G}^{\pm}(\zeta_{z}x_i^2,\mu)$ are perturbative kernels. From the rapidity evolution equation, one clearly see that as $P^z\rightarrow \infty$, there are large logarithms in $P^z$, part of it being non-perturbative and part of it being perturbative. Therefore, to match to the WF amplitude one needs a hard kernel $H$ to take into account the perturbative logarithms, and an exponential of $K_N$ to take into account the non-perturbative rapidity logarithms.

\subsection{Generalized Off-light-cone Soft Functions}

In the previous subsection, we have introduced the gauge-invariant quasi-WF amplitude. However, it still suffers from implicit scheme dependency since they are defined with an off-light-cone regulator along the $z$ direction. To match them to the physical WF amplitudes introduced in Sec.~\ref{Sec:LFWF}, one needs the (generalized) off-light-cone soft functions $S^{\pm}_N(\vec{b}_{i,\perp},\mu,Y,Y')$ composed of $N+1$ Wilson-line cusps which we now introduce.

We first  define the off-light-cone space-like vectors as $p\rightarrow p_Y= p-e^{-2Y}(P^+)^2n $, $n \rightarrow n_{Y'}=n-e^{-2Y'}\frac{p}{(P^+)^2}$ and the off-light-cone Wilson-line cusps ${\cal C}^{\pm}(b,Y,Y')$:
\begin{align}
&{\cal C}^{\pm}(\vec{b}_\perp,Y,Y')=W^{\pm}_{n_{Y'}}(\vec{b}_\perp)W^{\dagger}_{p_Y}(\vec{b}_\perp) \ ,
\end{align}
where the off-light-cone gauge-links $W_{p_Y}$ and $W_{n_Y'}$ are defined as
\begin{align}
&W_{p_Y}(\vec{b}_\perp)={\cal P}{\rm exp}\left[-ig\int_{0}^{-\infty} d\lambda' p_Y \cdot A(\lambda' p_Y+\vec{b}_\perp)\right] \ ,
\end{align}
and
\begin{align}
&W_{n_{Y'}}^{\pm}(\vec{b}_\perp)={\cal P}{\rm exp}\left[-ig\int_{0}^{\pm \infty} d\lambda  n_{Y'}\cdot A(\lambda n_{Y'}+\vec{b}_\perp)\right]  \ ,
\end{align}
respectively. With the off-light-cone Wilson-line cusps, the soft functions are defined in a way similar to Eq.~(\ref{eq:S_N}):
\begin{align}\label{eq:SNoff}
     &S^{\pm}_N(\vec{b}_{i\perp},\mu,Y,Y')\nonumber \\
     &= \frac{\langle 0|{\cal P}_N{\cal T}\prod_{i=0}^{N}{\cal C}^{\pm}(\vec{b}_{i\perp},Y,Y')|0\rangle}{\sqrt{Z_E(Y)}\sqrt{Z_E(Y')}}  \ ,
\end{align}
where $\sqrt{Z_E}$ is introduced similar as in Eq.~(\ref{eq:ZE}) to subtract the pinch-pole singularities and power divergences of the off-light-cone gauge-links. In terms of $\ln \rho^2=2(Y+Y')$, we can also write the off-light-cone soft functions as $S^{\pm}_N(\vec{b}_{i\perp},\mu,\rho)$.

In the light-cone limit $Y+Y'\rightarrow \infty$ , we have:
\begin{align}\label{eq:Sroff}
      &S^{\pm}_{N}(\vec{b}_{i\perp},\mu,Y,Y') \nonumber \\
      &={\rm exp}\left[ K_N(\vec{b}_{i\perp},\mu) \ln (\mp e^{Y+Y'}-i0) +{\cal D}_N(\vec{b}_{i\perp},\mu)\right]  \ ,
\end{align}
where $K_N$ is the same Collins-Soper kernel as in Eq.~(\ref{eq:SNon}), but ${\cal D}_N$ is different from the on-light-cone version ${\cal D}_{2,N}$.  Similar to the case of $\delta$ regulator, imaginary part appears in the $S_N^{+}$ case due to analyticity property. In fact, one can show that the the off-light-cone soft function depends only on the (complex) hyperbolic angle for the directions vectors from which the imaginary part can be generated. The rapidity-independent part is defined as the generalized reduced soft function:
\begin{align}
S_{rN}(b_{i\perp},\mu)=e^{-{\cal D}_N(\vec{b}_{i\perp},\mu)} \ ,
\end{align}
which is independent of the $\pm \infty$ choice. Based on the renormalization property of non-light-like Wilson-loops, the reduced soft function satisfies the RG equation
\begin{align}
   &\mu^2 \frac{d}{d\mu^2} \ln S_{rN}(\vec{b}_{i\perp},\mu)=-\frac{N+1}{2}\Gamma_S(\alpha_s)  \ ,
\end{align}
where $\Gamma_S$ is the constant part of the cusp-anomalous dimension at large hyperbolic cusp angle $Y+Y'$ for the off-light-cone soft function:
\begin{align}
&\mu^2 \frac{d}{d\mu^2} \ln S^{-}_{N}(\vec{b}_{i\perp},\mu,Y,Y')\nonumber \\
&\qquad\qquad=-(Y+Y')\frac{N+1}{2}\Gamma_{\rm cusp}(\alpha_s)+\frac{N+1}{2}\Gamma_S(\alpha_s) \ .
\end{align}
Notice that $\Gamma_S$ depends on the color representation ${\cal R}$.

Similar to the $N=1$ case~\cite{Ji:2019sxk}, the off-light-cone soft function $S^{-}_N(\vec{b}_{i\perp},\mu,Y,Y')$ equals to a time-independent form factor of fast-moving color-charged state. Thus it can be simulated using the Euclidean formalism of heavy-quark effective theory (HQET).
The explicit form of the HQET implementation depends on the color-representations of the Wilson-line cusps.
We will not go into the details here for general case.

In the next subsection, we will show that similar to the case of quasi-TMDPDFs~\cite{Ji:2019ewn}, the corresponding non-perturbative rapidity independent part that cancels the off-light-cone scheme dependencies in quasi-WF amplitudes is exactly the reduced soft function.

\subsection{Factorization of Quasi-WF Amplitudes}

Given the reduced soft function, we can state the matching formula between the quasi-WF amplitudes at finite momentum and that in LF theory :
\begin{align}\label{eq:quasiWF_fac}
&\widetilde \psi_N^{\pm}(x_i,\vec{b}_{i\perp},\mu,\zeta_{z})\sqrt{S_{rN}(\vec{b}_{i\perp},\mu)}=e^{\ln \frac{\mp\zeta_{z}-i0}{\zeta}K_N(\vec{b}_{i\perp},\mu)}\\ &\times H_N^{\pm}\left(\zeta_{z,i}/\mu^2\right)\psi_N^{\pm}(x_i,\vec{b}_{i\perp},\mu,\zeta)+... \ , \nonumber
\end{align}
where $ H_N^{\pm}\left(\zeta_{z,i}/\mu^2\right)$ is the perturbative matching kernel which is responsible for the large logarithms of $P^z$ generated by the perturbative ${\cal G}^{\pm}$ part of the momentum evolution equation. The $H_N^+$ and $H_N^-$ relates with each other through analytic continuation in $\zeta_z$ Similar to the quasi-TMDPDFs, the momentum fractions of the quasi-WF amplitudes and the LFWF amplitudes are the same since the momentum fractions can only be modified by collinear modes when $|\vec{k}_\perp|\ll P^z $. And $e^{\ln \frac{\mp\zeta_{z}-i0}{\zeta}K_N(\vec{b}_{i\perp},\mu)}$ is the part involving the non-perturbative rapidity evolution kernel. The imaginary part is required again in the $S_N^{+}$ case due to our definition of $S_{rN}$. As in the case of quasi-TMDPDFs, this factor is required to cancel the non-perturbative logarithms in $P^z$. The omitted terms are the power-corrections which are of order ${\cal O}\left(\Lambda^2_{\rm QCD}/\zeta_{z,i},M^2/\zeta_{z,i},1/(\delta \vec{b}_{ij,\perp}^2\zeta_{z,i}) \right)$ with $M$ being the hadron mass and $\delta \vec{b}_{ij,\perp}=\vec{b}_{i\perp}-\vec{b}_{j\perp}$.

\begin{figure}
\includegraphics[width=0.5\columnwidth]{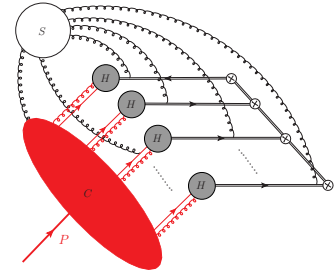}
\caption{\label{fig:WFreduced} The leading region for quasi-WF amplitude $\widetilde \psi^-_N$, where $C$ is the collinear sub-diagram, $S$ is the soft sub-diagram and $H$ are the hard sub-diagrams. The $N+1$ hard-cores are disconnected with each-other, therefore the momentum fractions of the quasi-WF amplitudes and physical WF amplitudes are the same. }
\end{figure}

Similar to the case of quasi-TMDPDF~\cite{Ji:2019ewn}, a sketch of mathematical proof for the factorization formula Eq.~(\ref{eq:quasiWF_fac}) can be provided with the help of the leading regions for quasi-LFWF amplitudes, shown in Fig.~\ref{fig:WFreduced}. There are collinear, soft and hard contributions captured by the corresponding collinear (C), soft (S) and hard (H) sub-diagrams. The collinear contributions are the same as the WF amplitudes defined with LF correlators, while the soft contribution can be factorized using off-light-cone soft functions. At large $P^z$ and large $|\vec{b}_{i\perp}-\vec{b}_{j\perp}|$, hard exchanges between vertices at different $\vec{b}_{i\perp}$ are power-suppressed, therefore the hard contributions are confined near the $N+1$ vertices. The hard natural scales for the hard sub-diagrams are given by $\zeta_{z,i}=\zeta_z x_i^2$, which are Lorentz-invariant combinations of the gauge-link direction $n^z$ and the parton's momenta $k_i=x_iP$. As a result, the momentum fractions $x_i$ only receive collinear contributions and remain the same between the quasi-WF amplitude and the physical LFWF amplitudes.

Given the leading region, a standard application of Ward-identity argument~\cite{Collins:2011zzd} will leads to the factorization formula Eq.~(\ref{eq:quasiWF_fac}). We use off-light-cone regulator in all the soft functions as in~\cite{Collins:2011zzd}. The reduced soft function $S_{rN}$ emerges at the final step as the non-cancelling soft function combination that appears when the factorization formula is expressed in terms of the physical LFWF amplitudes through the combination in Eq.~(\ref{eq:physicalWF}). Similar to the case of quasi-TMDPDFs, $S_{rN}$ serves to compensate the rapidity independent part between $\psi_N$ and $\tilde \psi_N$. The Collins-Soper kernel appears to compensate the rapidity mismatch between $\psi_N$ at $\zeta$ and $\tilde \psi_N$ at $\zeta_{z}$.

Here we should mention that when performing the factorization of quasi-WF amplitude using off-light-cone regulators, the emergence of $S_{rN}$ is expected since it is defined exactly through the off-light-cone soft functions. However, in the SCET style approach one can use on-light-cone regulators to do the factorization as well. As shown in Ref.~\cite{Ji:2020ect}, if one perform factorization of $\tilde \psi$ using on-light-cone regulators, the $S_{rN}$ can also be defined through the following type of combination
\begin{align} \label{eq:Sronlightcone}
    S_{rN}(\vec{b}_{i\perp},\mu)=\lim_{\delta^+,\delta^-\rightarrow 0}\frac{S_N^{-}(\vec{b}_{i\perp},\mu,\delta^+\delta^-)}{S_N^{-}(\vec{b}_{i\perp},\mu,\delta^+)S_N^{-}(\vec{b}_{i\perp},\mu,\delta^-)} \ ,
\end{align}
where the $S_N^{-}(\vec{b}_{i\perp},\mu,\delta^+\delta^-)$ in the numerator is the on-light-cone soft function defined in Eq.~(\ref{eq:SNon}).  The $S_N^{-}(\vec{b}_{i\perp},\mu,\delta^{\pm})$ in the denominator is defined similarly, but with one on-light-cone gauge-link direction along $p$ or $n$, and another off-light-cone one along $n^z$. The light-cone regulator for the on-light-cone gauge-link directions are denoted by $\delta^{\pm}$.  Notice that the combination in Eq.~(\ref{eq:Sronlightcone}) also appears in Ref.~\cite{Vladimirov:2020ofp} in the special case $N=1$ and is equivalent to their ``instant-jet TMD distribution'' $\Psi(\bar \xi)$ at special $\bar \xi$.

\subsection{Collins-Soper Kernel From Quasi-LFWF Amplitudes}

As an application of the factorization formula Eq.~(\ref{eq:quasiWF_fac}), we show that the Collins-Soper
rapidity evolution kernel can be calculated from the large-momentum dependence of the
quasi-WF amplitude, just like the case
of quasi-TMDPDF~\cite{Ji:2014hxa}.

Notice that by taking the ratio of two quasi-LFWFs, the reduced soft function $S_{rN}$ cancels. Furthermore, if we chose the quasi-LFWFs to have different sets of Collins-Soper scale, $\zeta_{z}$ and $\zeta'_{z}$ but with the same $x_i$, the LFWF amplitudes will also cancel.  This allows the Collins Soper kernel to be extract in the following way
\begin{align}
    &K_{N}(\vec{b}_{i\perp},\mu)\nonumber \\
    &=\frac{1}{\ln \frac{\zeta_{z}}{\zeta'_{z}}} \ln \frac{H_N^{\pm}\left(\zeta'_{z}/\mu^2\right)\widetilde \psi_N^{\pm}(x_i,\vec{b}_{i\perp},\mu,\zeta_{z})}{H_N^{\pm}\left(\zeta_{z}/\mu^2\right)\widetilde \psi_N^{\pm}(x_i,\vec{b}_{i\perp},\mu,\zeta'_{z})} \ .
\end{align}
It is clear that this is a result of the rapidity evolution equation for quasi-LFWFs. The same method was first obtained in the context of quasi-TMDPDF in~\cite{Ebert:2018gzl} by taking ratio of quasi-TMDPDFs at different $P^z$. Notice that the matching kernel and the quasi-LFWF amplitudes all have imaginary parts in general, but after taking the ratio, the imaginary parts cancels, left with a purely real $K_{N}$.

\section{Leading Wave Function Amplitude For Pseudo-scalar Meson }

In previous sections, we have presented the general theoretical framework to calculate LFWF amplitudes using LaMET formulations. As an application of the general principles, we present in this section
the example for the leading $\bar q q$ component wave function for a pseudo-scalar meson. As usual, we first present the standard definition in terms of the light-front formulation, then introduce the corresponding Euclidean formulation in large momentum expansion.

\subsection{The Light-Front Formulation}

\begin{figure}[!h]
\includegraphics[width=0.7\columnwidth]{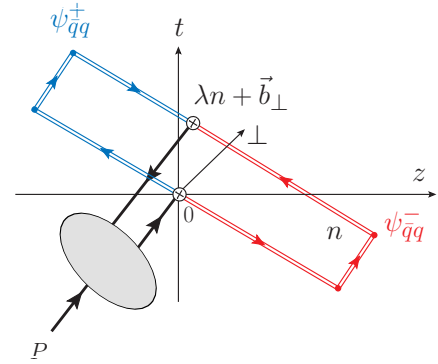}
\caption{\label{fig:WF2} The $\bar q q$ wave function for a pseudo-scalar meson. Again, the red case corresponds to $\psi_{\bar q q}^-$ and the blue case corrresponds to $\psi_{\bar qq}^+$.   }
\end{figure}

According to the general rules, the leading unsubtracted wave-function amplitude for a pseudo-scalar meson is defined by
\begin{align}
&\psi^{\pm 0}_{\bar q q}(x,b_\perp, \mu,\delta^-)=\frac{1}{2}\int \frac{d\lambda}{2\pi}e^{ix_r\lambda } \\ & \times \langle 0|\overline{\mit\Psi}_n^{\pm}(\lambda n/2+\vec{b}_\perp) \gamma^+ \gamma^5 {\mit\Psi}_n^{\pm}(-\lambda n/2)|P\rangle\Big|_{\delta^-}  \ , \nonumber
\end{align}
where $x_r$ is related to the standard definition of $x$ by $x_r=x-1/2$, and the ``gauge-invariant'' quark field is
\begin{align}
   {\mit \Psi}^{\pm}_n(\xi) = W^{\pm }_n(\xi)|_{\delta^-}\psi(\xi) \ .
\end{align}
 Due to rotational invariance, the amplitude defined above is a function of $b_\perp=|\vec{b}_\perp|$, thus we have omitted the vector arrow for $\vec{b}_\perp$, and we will do so throughout the section for the $N=1$ case. See Fig.~\ref{fig:WF2} for a depiction of the meson wave functions.

We now present the perturbative
one-loop calculation for the above amplitude. We consider a system where the incoming quark and anti-quark momenta are $x_0P^+$ and $(1- x_0) P^+$, respectively. The spin projection operator for the incoming state is proportional to $\gamma^5 \gamma^-$ and the tree-level wave-function amplitude is normalized to $\delta\left(x-x_0\right)$. Evaluated in the $\delta$ regularization scheme, the bare WF amplitude reads
\begin{align}
&\psi^{\pm0}_{\bar q q}(x,b_\perp, \mu,\delta^-)\nonumber \\
&=\frac{\alpha_s C_F}{2\pi}\left[F(x,x_0,b_\perp,\mu)\right]_{+}+\frac{\alpha_s C_F}{2\pi}\delta\left(x-x_0\right)\nonumber \\
&\times \Bigg\{ L_b\left(\frac{3}{2}+\ln \frac{-(\delta^-)^2\mp i0}{x_0\bar x_0}\right)+\frac{1}{2} \Bigg\} \ ,
\end{align}
where $C_F=(N_c^2-1)/2N_c$ and $L_b=\ln \frac{\mu^2b^2_\perp}{4e^{-2\gamma_E}}$ and
\begin{align}
&F(x,x_0,b_\perp,\mu)\nonumber \\
&=\left[-\left(\frac{1}{\epsilon_{\rm IR}}+L_b\right)\left(\frac{x}{x_0(x_0-x)}+\frac{x}{x_0}\right)+\frac{x}{x_0}\right]\nonumber \\
&\times \theta(x)\theta(x_0-x)+\left(x\rightarrow \bar x,x_0\rightarrow \bar x_0
\right) \ ,
\end{align}
where $\frac{1}{\epsilon_{\rm IR}}$ indicates that there is an IR divergence. Notice that $x$ and $\bar x=1-x$ are the momentum fractions carried by the quark and the anti-quark.

The soft function with $N=1$ and ${\cal R}=\{f,\bar f \}$ is defined with two Wilson-line cusps explicitly as
\begin{align}
&S^{\pm}_1(b_\perp,\mu,\delta^+\delta^-)=\frac{1}{N_c} {\rm tr}\langle0|{\cal T}W^{-\dagger}_{p}(b)|_{\delta^+}\nonumber \\ & \times W^{\pm}_{n}(b)|_{\delta^-} W^{\pm \dagger}_n(0)|_{\delta^-} W^-_p(0)|_{\delta^+}|0\rangle \ .
\end{align}
At one-loop, the soft function $S^{\pm}_{1}(b_\perp,\mu,\delta^+,\delta^-)$ is given by~\cite{Echevarria:2012js} :
\begin{align}
&S^{-}_{1}(b_\perp,\mu,\delta^+\delta^-)\nonumber \\ &=1+\frac{\alpha_s C_F}{2\pi}\left(L_b^2-2L_b \ln \frac{\mu^2}{2(P^+)^2\delta^+\delta^-}+\frac{\pi^2}{6}\right) \ , \\
&S^{+}_{1}(b_\perp,\mu,\delta^+\delta^-)\nonumber \\ &=1+\frac{\alpha_s C_F}{2\pi}\bigg(L_b^2-2L_b \left(\ln \frac{\mu^2}{2(P^+)^2\delta^+\delta^-}+i\pi\right)+\frac{\pi^2}{6}\bigg) \ .
\end{align}
where $S^+$ contains an imaginary part.
Therefore, we can extract at the leading order the CS kernel and reduced soft function,
\begin{align}
K_1(b_\perp,\mu)&=-\frac{\alpha_sC_F}{\pi} L_b  \,,\\
{\cal D}_{2,1}(b_\perp,\mu)&=\frac{\alpha_s C_F}{2\pi}\left(L_b^2+\frac{\pi^2}{6}\right) \ .
\end{align}
They are consistent with the case for TMDPDFs. The rapidity dependence coming from the initial-state quark radiation is intrinsic and nonperturbative for large $b_\perp$.

In term of these, the renormalized WF amplitude is defined explicitly as
\begin{align}
\psi^{\pm}_{\bar q q}(x,b_\perp,\mu,\zeta)=\lim_{\delta^-\rightarrow 0}\frac{\psi^{\pm0}_{\bar qq}(x,b_\perp,\mu,\delta^-)}{\sqrt{S^{\pm}_1(b_\perp,\mu,\delta^-e^{2y_n},\delta^-)}} \ ,
\end{align}
where we have chosen $\delta^+ = e^{2y_n}\delta-$ with $y_n$
as a dimensionless parameter.
While both $\psi^0$ and $S_1$ depend on the regulator $\delta^{\pm}$, the combination $\psi$ is regularization independent and gives rise to the dependencies on the universal rapidity variables $\zeta=2(xP^+)^2e^{2y_n}$ and $\bar\zeta=2(\bar x P^+)e^{2y_n}$, with the dependence on the latter being omitted. Combining the results above, the one-loop WF amplitude reads
\begin{align}
&\psi^{\pm}_{\bar q q}(x,b_\perp, \mu,\zeta) \\
&=\frac{\alpha_s C_F}{2\pi}\left[F(x,x_0,b_\perp,\mu)\right]_{+}+\frac{\alpha_s C_F}{2\pi}\delta\left(x-x_0\right)\nonumber \\
&\times \Bigg\{-\frac{L_b^2}{2}+L_b\left(\frac{3}{2}+\ln \frac{\mu^2}{\pm \sqrt{\zeta \bar \zeta}-i0}\right)+\frac{1}{2}-\frac{\pi^2}{12}\Bigg\} \ , \nonumber
\end{align}
which effectively replaces the rapidity regulator $\delta$ by the rapidity scale $\zeta$. It is important to note that the above result is now independent of the light-cone regulator $\delta$.

The renormalized WF amplitude satisfies the rapidity (momentum) evolution equation
\begin{align}
2\zeta \frac{d}{d\zeta} \ln  \psi^{\pm}_{\bar q q}(x,b_\perp, \mu,\zeta)=K_1(b_\perp,\mu) \ ,
\end{align}
and the RGE:
\begin{align}
&\mu^2\frac{d}{d\mu^2}\ln \psi^{\pm}_{\bar q q}(x,b_\perp, \mu,\zeta)\nonumber \\
&=\frac{1}{2}\Gamma_{\rm cusp}(\alpha_s)\ln \frac{\mu^2}{\pm \sqrt{\zeta \bar \zeta}-i0}-\gamma_{H}(\alpha_s) \ .
\end{align}
In the above equations, the evolution kernel $K_1(b_\perp, \mu)$ and the anomalous dimensions are the same as those of
the TMDDPFs.  At one-loop, the above cusp and hard anomalous dimensions read
\begin{align}\label{eq:ad_hard}
\Gamma_{\rm cusp}(\alpha_s) =\frac{\alpha_s C_F}{\pi}; ~~~~~
\gamma_H(\alpha_s) =-\frac{3\alpha_sC_F}{4\pi}  \ .
\end{align}
Recently the cusp anomalous dimension have been calculated to 4-loops~\cite{Henn:2019swt,vonManteuffel:2020vjv}.

\subsection{LaMET Expansion}

\begin{figure}[!h]
\includegraphics[width=0.7\columnwidth]{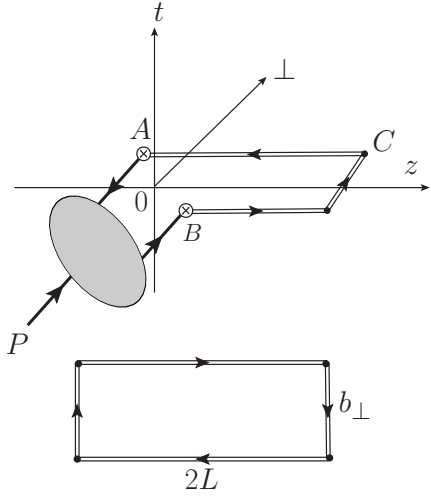}
\caption{\label{fig:quasi2} The quasi-LFWF $\tilde \psi^-_{\bar qq}$ (upper) and the $Z_E$ (lower). In the figure, $A=\lambda n_z/2+\vec{b}_\perp/2$, $B=-\lambda n_z/2-\vec{b}_\perp/2$ and $C=Ln_z+{\vec b}_\perp$. The crosses denote the quark-link vertices. }
\end{figure}

To calculate the above $q\bar q$ WF amplitude, we define the un-subtracted quasi-WF amplitude as
\begin{align}\label{eq:quasiWF1}
&\widetilde \psi_{\bar q q}^{\pm }(x,b_\perp,\mu,\zeta_z)\\
&=\!\lim_{L\to\infty}\int \frac{d\lambda}{4\pi}e^{-i x_r \lambda}\frac{\langle 0|\overline{\mit\Psi}_{\mp {n_z}}(\frac{\lambda {n_z}}{2}\!+\!\vec{b}_\perp)\Gamma {\mit\Psi}_{\mp {n_z}}(\!-\!\frac{\lambda {n_z}}{2})|PS\rangle}{\sqrt{Z_E(2L,b_\perp,\mu)}} \ ,\nonumber
\end{align}
where $x_r=x-\frac{1}{2}$ and ${\mit \Psi}_{\mp {n_z}}$ now contains a gauge-link along the $\mp z$ direction pointing to $\mp L {n_z}$, similar to Eq.~( \ref{eq:invariantnz}). The Wilson-line self-energy and interaction
are subtracted, which is also similar to the case for quasi-TMDPDF. See Fig.~\ref{fig:quasi2} for a depiction of the quasi-LFWF amplitudes $\psi_{\bar q q}^{-}$ and $Z_E$ for pseudo-scalar meson. Note that $\widetilde \psi$ depends on $\zeta_z=(2xP\cdot {n_z})^2$, $\bar \zeta_z=(2\bar xP\cdot {n_z})^2$ and the renormalization scale $\mu$. More generally, the quasi-WF amplitude satisfies the renormalization group equation
\begin{align}
\mu^2 \frac{d}{d\mu^2}\ln \widetilde \psi^{\pm}_{\bar q q}(x,b_\perp,\mu,\zeta_z)=\gamma_{F}(\alpha_s) \ ,
\end{align}
where $\gamma_F$ is the anomalous dimension for a heavy-light current. This is due to the fact that the quasi-WF amplitude, after the self-energy and corner-divergence subtraction contains only logarithmic UV divergences associated with quark-link vertices.
\begin{figure}[!h]
\includegraphics[width=1\columnwidth]{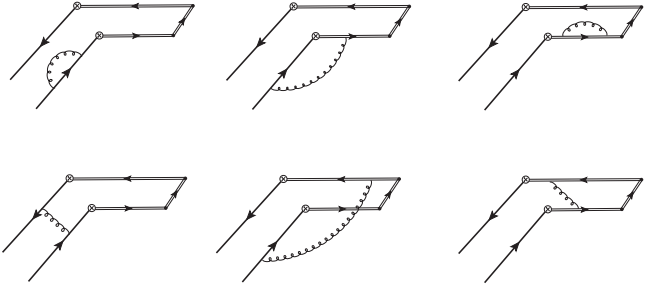}
\caption{\label{fig:quasione} One-loop diagrams for mesonic LFWF amplitudes and quasi-LFWF amplitudes. The meson state is treated as a pair of free quark and antiquark. The first line represents ``virtual'' diagram and the second line represents ``real'' diagram. }
\end{figure}
The one-loop quasi-WF amplitude receives contribution from more diagrams compared to the WF amplitude. Unlike the LFWF amplitude, all the ``virtual'' diagrams and gauge-link self-interactions are non-vanishing. The total result reads
\begin{align}\label{eq:quasiWF1oop}
&\widetilde \psi^{\pm}_{\bar q q}(x,b_\perp, \mu,\zeta_z) \\&
=\frac{\alpha_s C_F}{2\pi}\left[F(x,x_0,b_\perp,\mu)\right]_{+}+\frac{\alpha_s C_F}{2\pi}\delta\left(x-x_0\right)\nonumber \\
&\times \Bigg\{-\frac{L_b^2}{2}+L_b\left[\frac{5}{2}+\ln \frac{\mu^2}{-\sqrt{\zeta_z \bar \zeta_z} \pm i0}\right]-\frac{3}{2}-\frac{\pi^2}{2} \nonumber \\
& +\left[-\frac{1}{4}\ln^2\frac{-\zeta_z\pm i0}{\mu^2}+\frac{1}{2}\ln \frac{-\zeta_z\pm i0}{\mu^2}+(\zeta_z \rightarrow \bar \zeta_z)\right]\Bigg\}\ . \nonumber
\end{align}
The imaginary parts are all caused by the rapidity logarithms in terms of $\frac{(2xP\cdot {n_z})^2}{{n_z}^2}=-(2xP^z)^2$, and the proper $i0$ choices are again determined by the $i0$ prescriptions in the gauge-link propagators that guarantee exponential decay at infinities. The results here are consistent with the off-light-cone WF amplitudes calculated in Ref.~\cite{Ma:2006dp} through analytic continuation where gauge-links are chosen to be time-like.

The one-loop off-light-cone soft function reads ~\cite{Ebert:2019okf},
\begin{align}
&S^{-(1)}_{1}(b_\perp,\mu,Y,Y')=\frac{\alpha_sC_F}{\pi}\big[1-(Y+Y')\big]L_b \ , \\
&S^{+(1)}_{1}(b_\perp,\mu,Y,Y')=\frac{\alpha_sC_F}{\pi}\big[1-i\pi-(Y+Y')\big]L_b \ ,
\end{align}
from which the reduced soft function can be extracted as
\begin{align}
    S^{(1)}_{r1}(b_\perp,\mu)=-\frac{\alpha_sC_F}{\pi}L_b \ .
\end{align}
These results are the same as the case of TMDPDFs.  From these one can extract the one-loop value of $\Gamma_S$ as $\Gamma_S^{(1)}=\alpha_s C_F/\pi$ , while and at two-loop level one has
\begin{align}
\Gamma_S^{(2)}=\frac{\alpha_s^2}{\pi^2}\Big[C_FC_A\big(-\frac{49}{36}+\frac{\pi^2}{12}-\frac{\zeta_3}{2}\big)+C_FN_F\frac{5}{18}\Big] \ ,
\end{align}
which can be extracted from the generic results in~\cite{Grozin:2015kna}.

\begin{figure}[!h]
\includegraphics[width=0.7\columnwidth]{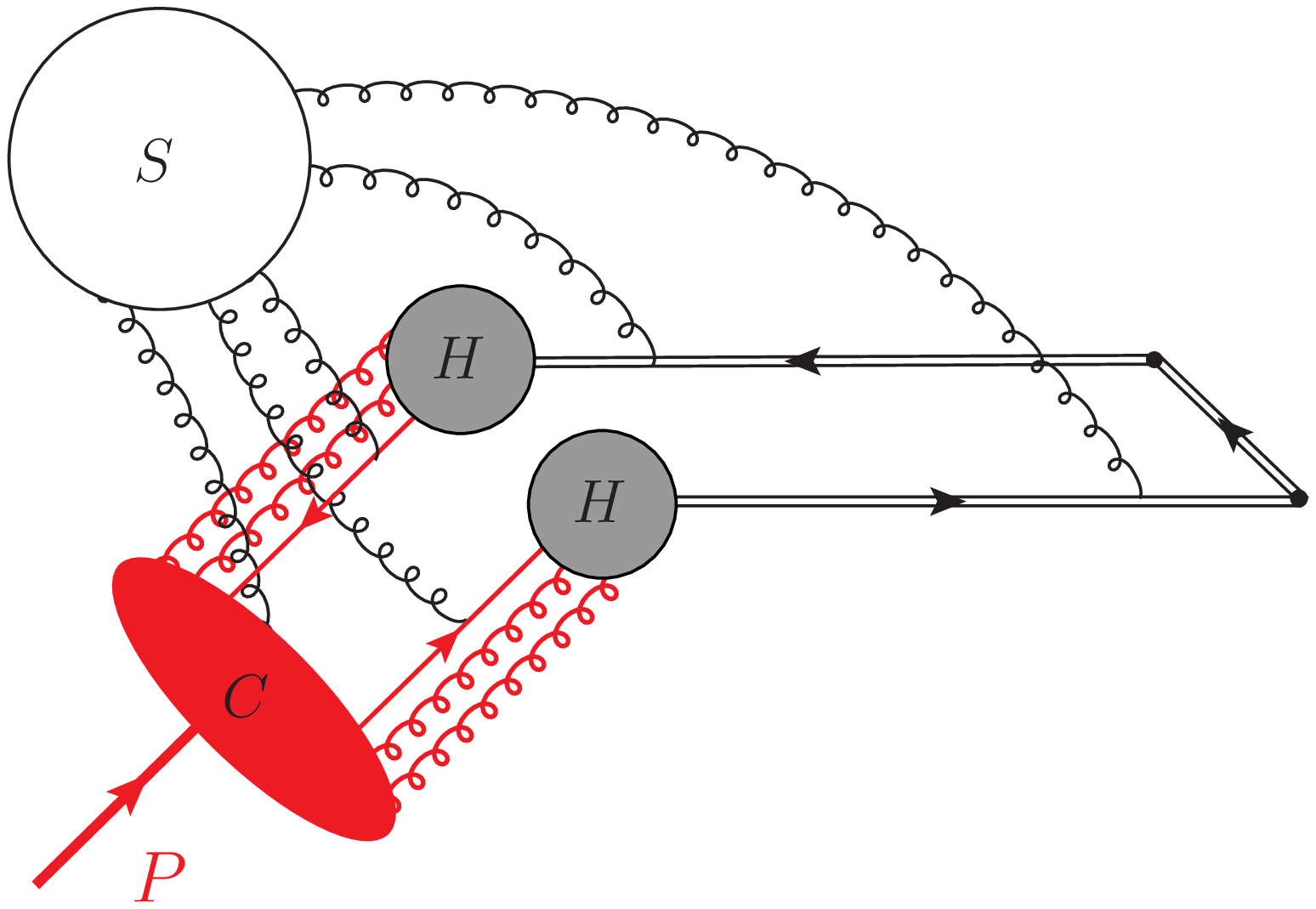}
\caption{\label{fig:WF2reduced} The leading region for quasi-WF amplitude $\widetilde \psi^-_{\bar q q}$ for a ppseudo-scalar meson, where $C$ is the collinear sub-diagram, $S$ is the soft sub-diagram and $H$ are the hard sub-diagrams. Similar to the general cases, the two hard-cores are disconnected with each-other,therefore the momentum fractions of the quasi-WF amplitudes and physical WF amplitudes are the same. }
\end{figure}

The matching formula between the quasi-LFWF amplitude and the light-front one at large $P_z$ is:
\begin{align}\label{eq:quai_fac_1}
&\widetilde \psi^{\pm}_{\bar q q}(x,b_\perp,\mu,\zeta_z)S_{r1}^{\frac{1}{2}}(b_\perp,\mu) \\
&=H^\pm_1\left(\zeta_z/\mu^2,\bar \zeta_z/\mu^2\right)e^{\ln \frac{\mp \zeta_z-i0}{\zeta}K(b_\perp,\mu)}\psi_{\bar qq}^\pm(x,b_\perp,\mu,\zeta) \ , \nonumber
\end{align}
where $H_1$ is the perturbative matching kernel. The physics reason for this factorization formula is similar to that for the TMDPDF: Our un-subtracted quasi-WF amplitude is defined the off-light-cone scheme. By comparing the
TMD factorization in both off and on-the-light-cone schemes, one obtains the matching formula above. See Fig.~\ref{fig:WF2reduced} for a depiction of the corresponding leading region.

Combining the RGEs for the WF amplitude, the reduced soft function, and the quasi-WF amplitude, the matching kernel satisfies a simple renormalization group equation:
\begin{align}
&\mu^2\frac{d}{d\mu^2}\ln H_1^{\pm}\left(\zeta_z/\mu^2,\bar \zeta_z/\mu^2\right)\nonumber \\
&=\frac{1}{2}\Gamma_{\rm cusp}(\alpha_s) \ln \frac{-\sqrt{\zeta_z\bar \zeta_z}\pm i0}{\mu^2} + \frac{1}{2}\gamma_C(\alpha_s)\ ,
\end{align}
where $\gamma_C(\alpha_s)=2\gamma_F(\alpha_s)- \Gamma_S(\alpha_s) +2\gamma_H(\alpha_s) $ with $\gamma_F(\alpha_s)$ the anomalous dimension for heavy-light current and $\Gamma_S(\alpha_s)$ the constant part for the cusp anomalous dimension at large cusp angle, and $\gamma_H(\alpha_s)$ the hard-anomalous dimension.

It is convenient to write the matching kernel in the exponential form, $H=e^{h}$. At one-loop level, $h$ can be extracted as:
\begin{align}
& h_1^{\pm(1)}\left(\zeta_z/\mu^2,\bar \zeta_z/\mu^2\right)\nonumber \\
&=\alpha_s\Bigg\{ c_1+\frac{C_F}{4\pi}\left[\ell_{\pm}+\bar \ell_{\pm}-\frac{1}{2}(\ell_{\pm}^2+\bar \ell_{\pm}^2)\right]\Bigg\} \ .
\end{align}
At two-loop level, we anticipate
\begin{align}
& h_1^{\pm(2)}\left(\zeta_z/\mu^2,\bar \zeta_z/\mu^2\right)\\
&=\alpha_s^2 c_2-\frac{1}{4}\left[\gamma^{(2)}_C-\alpha_s^2\beta_0 c_1\right](\ell_{\pm}+\bar \ell_{\pm})\nonumber\\
&-\frac{1}{8}\left[\Gamma^{(2)}_{\rm cusp}-\frac{\alpha_s^2\beta_0 C_F}{2\pi}\right](\ell_{\pm}^2+\bar \ell_{\pm}^2)-\frac{\alpha_s^2\beta_0C_F}{48\pi}(\ell_{\pm}^3 +\bar \ell_{\pm}^3)\nonumber \ ,
\end{align}
where we have $\ell_{\pm}=\ln \frac{-\zeta_z\pm i0}{\mu^2}$ and $\bar \ell_{\pm}=\ln \frac{-\bar \zeta_z\pm i0}{\mu^2}$. $\beta_0=-\left(\frac{11}{3}C_A-\frac{4}{3}N_f T_F\right)/(2\pi)$ is the coefficient of one-loop $\beta$-function,  $c_1=\frac{C_F}{2\pi}\left(-\frac{5\pi^2}{12}-2\right)$ and $c_2$ are constants.

The generalization to the lowest Fock component of the nucleon state can be done similarly~\cite{Ji:2002xn}. The WF amplitudes depend on two transverse separations $\vec{b}_{1\perp}$, $\vec{b}_{2\perp}$ and three momentum fractions $x_1+x_2+x_3=1$. And the soft function $S_2$ in each light-cone direction now consists of three gauge-links in fundamental representation, piecing together by the ${\rm SU}$(3) invariant tensor $\epsilon^{ijk}$. Previous discussions on the nucleon form factors and wave-functions can be found in~\cite{Lepage:1979za,Duncan:1979hi,Aznaurian:1979zz,Sterman:1997sx,Li:1992ce}. The nucleon
WF amplitude has also been discussed and calculated in various LF phenomenology models~\cite{Mondal:2020mpv,Du:2019qsz}. The LaMET formalism allows a first-principle determination of the amplitude following the general procedures discussed above.

\section{Conclusion}

In this paper, we present the LaMET formulation for computing LFWF amplitudes in QCD. We first reviewed the LFQ approach to QCD and its conceptual difficulties, especially the LF divergence related to small $k^+$ modes. These difficulties signals that the QFT on the LF must be viewed as an effective theory in which small $k^+$ modes are ``integrated out''. The LaMET provides a ``two-step'' approach to the LFQ physics and achieves the goal of LFQ without performing the LFQ explicitly.

We then formulate the LFWF amplitudes as gauge-invariant LF correlators. To maintain gauge-invariance, light-like gauge-links extending to infinities are required in their definitions and leads to rapidity divergences which must be regulated through rapidity regulators. The naive LFWF amplitudes depends on the non-physical rapidity regulator, which must be removed with the help of generalized soft functions composed of arbitrary numbers of Wilson-line cusps. By combining the un-subtracted WF amplitudes with square-root of generalized soft functions, one can construct the scheme-independent WF amplitudes that can be used in factorization formulas for physical quantities. Similar to the TMDPDFs, the physical WF amplitudes depends on additional rapidity scales the evolution of which is controlled by the CS kernels. Results for the $q\bar q$ wave function for a pseudo-scalar meson are presented in detail as an application of the general principles.

After introducing a standard version of LFWF amplitudes, we start to present their LaMET formulation. We carefully define the quasi-LF amplitudes in which operator and gauge-links are all time-independent. The large hadron momentum $P^z$ plays the role of a physical off-light-cone regulator. At large $P^z$, the quasi-LFWF amplitudes can be matched to the physical LFWF amplitudes by subtracting out the off-light-cone scheme dependency with the help of reduced soft functions. We introduce the generalized off-light-cone soft function and their properties, and study the factorization of quasi-WF amplitudes. The factorization formula is presented and supported by a sketch of its proof. As an application of the factorization formula, we show that the Collins-Soper kernels can be extracted from ratio of quasi-WF amplitudes in which the soft function contribution cancels.  Again, results for the pseudo-scalar meson wave functions are presented in details, including a prediction of two-loop matching kernel.

{\it Acknowledgment.}---We thank M. Burkhardt, W. Wang, Feng Yuan, Jianhui Zhang and Yong Zhao for valuable discussions. This work is supported partially by the US DOE, Office of Science, grant DE-SC0020682.

\bibliography{bibliography}

\end{document}